\documentclass[twocolumn]{aastex61}
\shorttitle{Microlensed Fe K$\alpha$ X-Ray Emission}
\shortauthors{Krawczynski \& Chartas}

\newcommand{\simgt}{\lower 2pt \hbox{$\, \buildrel {\scriptstyle >}\over {\scriptstyle\sim}\,$}}
\newcommand{\simlt}{\lower 2pt \hbox{$\, \buildrel {\scriptstyle <}\over {\scriptstyle\sim}\,$}}

\begin{document}


\title{Simulations of the Fe K$\alpha$ Energy Spectra from Gravitationally Microlensed Quasars}


\correspondingauthor{Henric Krawczynski}
\email{krawcz@wustl.edu}
\author{H. Krawczynski}
\affil{Physics Department and McDonnell Center for the Space Sciences, Washington University in St. Louis,
1 Brookings Drive, CB 1105, St. Louis, MO 63130, USA}
\author{G. Chartas}
\affil{Department of Physics and Astronomy, College of Charleston, Charleston, SC, 29424, USA}
\affil{Department of Physics and Astronomy, University of South Carolina, Columbia, SC, 29208}
\begin{abstract}
The analysis of the {\it Chandra} X-ray observations of the gravitationally lensed quasar 
RX~J1131$-$1231 revealed the detection of multiple and energy-variable spectral peaks.
The spectral variability is thought to result from the microlensing of the Fe K$\alpha$ emission, 
selectively amplifying the emission from certain regions of the accretion disk with certain effective frequency shifts of the 
Fe K$\alpha$ line emission.
 In this paper, we combine detailed simulations of the emission of Fe K$\alpha$ photons from the accretion disk 
of a Kerr black hole with calculations of the effect of gravitational microlensing on the observed energy spectra. 
The simulations show that microlensing can indeed produce multiply peaked energy spectra.
We explore the dependence of the spectral characteristics on black hole spin, 
accretion disk inclination, corona height, and microlensing amplification factor, 
and show that the measurements can be used to constrain these parameters. 
We find that the range of observed spectral peak energies of QSO RX~J1131$-$1231 can only 
be reproduced for black hole inclinations exceeding 70$^{\circ}$ and for 
lamppost corona heights of less than 30 gravitational radii above the black hole. 
We conclude by emphasizing the scientific potential of studies of the microlensed 
Fe K$\alpha$ quasar emission and the need for more detailed modeling that explores 
how the results change for more realistic accretion disk and corona geometries and
microlensing magnification patterns. A full analysis should furthermore model 
the signal-to-noise ratio of the observations and the resulting detection biases.
\end{abstract}
\keywords{accretion, accretion disks, black hole physics, gravitational lensing: strong, gravitational lensing: micro,
line: formation, line: profiles, 
(galaxies:) quasars: emission lines, 
(galaxies:) quasars: general,
(galaxies:) quasars: individual (QSO RX~J1131$-$1231), 
(galaxies:) quasars: supermassive black holes 
 }


\section{Introduction \label{sec:intro}}
	Observations of gravitationally lensed Quasi-Stellar Objects (QSOs) have emerged as a powerful tool 
to constrain the inner structure of QSO accretion flows. Optical, UV and X-ray observations of several
QSOs have revealed  the presence of microlensing, i.e. variations of the brightness of individual macro-images 
caused by the relative movement of the observer, the lensing galaxy, and the observed source. 
\citet{2010ApJ...712.1129M} use microlensing observations in the optical and UV to constrain 
the 2500~\AA~half-light radii of the accretion disks of eleven gravitationally lensed quasars. 
Although the inferred half-light radii scale with the black hole mass, they exceed the values 
predicted by thin disk theory by a factor of three.  While the optical/UV accretion disks sizes are thus
surprisingly large, {\it Chandra} observations of the microlensed X-ray emission from a sample of QSOs 
constrain the X-ray bright regions to be extremely compact. The observations of the quasars 
HE~1104$-$1805 ($z_{\rm S}$=2.31, $z_{\rm L}$=0.73) 
SDSS~0924$+$0219 ($z_{\rm S}$=1.52, $z_{\rm L}$=0.393), and 
RX~J1131$-$1231 ($z_{\rm S}$=0.658, $z_{\rm L}$=0.295) constrain the X-ray half-light radii to be smaller 
than 30 $r_{\rm g}$ with $r_{\rm g}=GM/c^2$ being the gravitational radius of the black hole
\citep{Char:09,Dai:10,Morg:12,Mosq:13,Blac:15,MacL:15}.
The microlensing constraints on the corona sizes of QSOs are independent, complimentary, and consistent
to the spectral  \citep[e.g.][]{Wilm:01,Fabi:09,Zogh:10,Park:14,Chia:15} and reverberation 
\citep[e.g.][]{Zogh:12,Cack:14,Uttl:14} 
constraints on the corona sizes of  nearby Narrow Line Seyfert I (NRLS I).

In this paper we focus on the observations of energy dependent microlensing reported by \citet{Char:12,Char:16,Char:17}. Monitoring a sample of eight gravitationally lensed quasars with {\it Chandra}, the team discovered three 
sources with multiple energy and variable emission lines features. For the best studied source, 
RX~J1131$-$1231, the authors analyzed 38 {\it Chandra} observations acquired 
between 4/12/2004 and 7/12/2014,  and detect 78 lines out of 152 energy 
spectra (one energy spectrum for each of the four quasar images for each of the 38 observations) on a 
$>$90\% confidence level.
The authors explain the discovery as resulting from the microlensing of the Fe~K$\alpha$ emission from 
the inner accretion flow. The Fe~K$\alpha$ X-rays are thought to originate from 
reprocessing of the coronal X-rays in the accretion disk as Fe~K$\alpha$ fluorescent photons 
 \citep[e.g.][and references therein]{2014SSRv..183..277R}.
The gravitational lensing by the stars of the lensing galaxy creates a complex network of caustics -- source plane locations where the magnification (flux amplification) of the light from a point-like source diverges. The movement of the caustics across the accretion disk owing to the relative movement of the source, the lens, and the observer can lead to time variable energy spectra, as the caustics selectively magnify the emission from small regions of the accretion disk with certain net Doppler and gravitational frequency shifts.
Here and in the following we refer to the frequency shift between the emission 
and observation of an Fe~K$\alpha$ photon as the $g$-factor:
\begin{equation}
g\,=\,\frac
{u_{\mu}k^{\mu}|_{\rm obs}}
{u_{\mu}k^{\mu}|_{\rm em}}
\end{equation} 
where $u^{\mu}$ is the four velocity of the observer (numerator) or emitting plasma (denominator)
and $k^{\mu}$ is the photon's wave vector in the respective reference frame.
Note that $g$ depends on the frequency shift incurred after emission (and thus on the motion of the emitting plasma), 
during the propagation through the curved black hole spacetime
(and thus on the spin and inclination of the black hole), and upon detection of the photon by the observer, 
and will be derived in the following with the help of the ray tracing code.

\citet[][C17]{Char:17} presented analytical estimates of the minimum and maximum g-factors of the Fe K$\alpha$ emission as function of black hole spin, the inner radius of the accretion disk, and the black hole inclination. The analytical estimates were used to constrain the inclination of the accretion disk to $i > 75^{\circ}$ and the inner accretion disk edge to 
$r< 8 r_{\rm g}$. The paper included first results from numerical ray tracing simulations supporting these constraints. 

Our numerical simulations model the illumination of the accretion disk with hard corona X-rays inducing the emission of Fe K$\alpha$ fluorescent photons and the propagation of the photons from their point of origin to a distant observer through the Kerr space time of the spinning black hole. Furthermore, they account for the magnification effects caused by the gravitational macro and microlensing. The numerical calculations improve over the analytical estimates in two ways: they allow us to correctly calculate the net gravitational and Doppler frequency shift which depends on the emission angle of the photons reaching the observer. The angles can only be determined by numerically solving the geodesic equation. 
Furthermore, the numerical simulations reveal the net intensity of the emission from certain parts of the accretion disk and thus indicate whether microlensed emission can indeed produce shifted and distorted peaks in the observed energy spectra. 

In this paper, we provide the details of the numerical simulations that were presented in the C17 paper. Furthermore, we present simulated results for a broad range of black hole spins, black hole inclinations, corona heights, and microlensing amplification factors, rather than for a single combination.    
We use the results to evaluate if observations can in principle constrain these parameters.

Compared to the earlier study of \citet{Popo:03}, we use the lamppost model \citep{1991A&A...247...25M}
to simulate the Fe K$\alpha$ emission (rather than assuming a power law emissivity profile 
in the radial Boyer Lindquist coordinate $r$) and we systematically study the dependence 
of the observed line properties on black hole spin and inclination and the lamppost properties.
Although a point-like corona is clearly a simplification, it seems to describe spectral \citep[e.g.][and references therein]{2014SSRv..183..277R}
and reverberation observations of nearby bright Seyfert 1 galaxies \citep[e.g.][and references therein]{Uttl:14} rather well.

We ran the simulations having in mind their applicability to the QSO RX~J1131$-$1231.
The source harbors a supermassive black hole with a mass between $8\times10^7 M_{\odot}$ and $2\times10^8 M_{\odot}$ \citep{2012A&A...544A..62S} accreting at
between a few percent and a few ten percent of the Eddington luminosity.
Mass accretion at this rate is believed to proceed via geometrically thin, optically thick 
accretion disks \citep{1973A&A....24..337S,1973blho.conf..343N,1974ApJ...191..499P,
Nara:05,2014MNRAS.441.3177M}.
As we limit the analysis to the reflected Fe~K$\alpha$ emission from the accretion disk 
irradiated by a lamppost corona, our results are largely independent of the detailed properties 
of the accretion disk itself as long as the disk is geometrically thin
(see the discussion of the impact of different absorption and 
Fe K$\alpha$ emission efficiencies below).
Note that recent relativistic (radiation) magnetohydrodynamical simulations indicate 
that the analytical description of the thin disks is rather accurate
\citep{2011ApJ...743..115N,2011MNRAS.414.1183K,2012MNRAS.420..684P,2016MNRAS.459.4397S}. 
We model the microlensing with a simple parameterization of the magnification close to caustic folds.
The results obtained with more realistic magnification maps (including curved folds, cusps, and 
combinations of these), based on the inverse ray shooting method and a systematic exploration of the
microlensing parameter space, and actual fits of the simulated data to {\it Chandra} data will be 
presented in forthcoming publications.

The rest of the paper is structured as follows: we describe the numerical methods in \S\ref{methods}, 
including the general relativistic ray tracing code and the modeling of the microlensing 
magnification. The results of the simulations are described in \S\ref{results}. 
We analyze the constraints on the black hole, accretion disk, corona, and microlensing parameters 
that can be derived from the observed spectral peak properties in \S\ref{ML}. We summarize 
and discuss the results in \S\ref{summary}. 
\section{Methodology}
\label{methods}
\subsection{Method of combining the ray tracing simulations of the Fe~K$\alpha$ emission with the magnification maps}
\begin{figure}
\includegraphics[angle=270,width=0.47\textwidth]{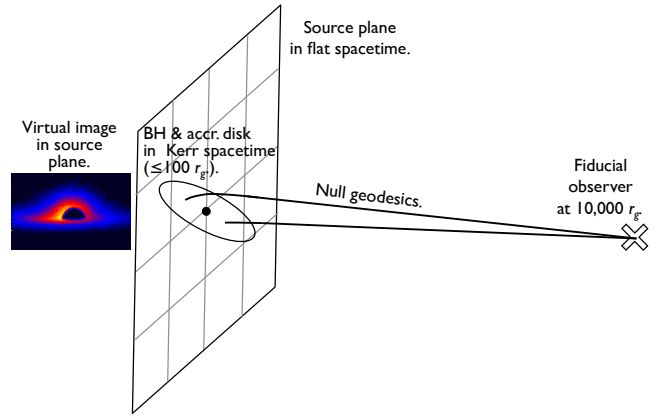}
\caption{\label{fig:sketch} We perform general relativistic ray tracing simulations in the Kerr spacetime of photon packets 
from a lamppost corona illuminating an accretion disk extending from $r=r_{\rm ISCO}$ to $r\,=\,100\,r_{\rm g}$ 
and prompting the emission of Fe~K$\alpha$ photon packets (left side of diagram). 
The latter are tracked until they reach a coordinate stationary observer at $r_{\rm obs}\,=\,10,000\,r_{\rm g}$
(right side of diagram). Assuming a flat Minkowskian space time, we back-project each photon packet 
to a source plane at $10,000\,r_{\rm g}$ from the observer based on its wavevector $\check{k}^{\mu}$ 
in the observer's reference frame. The rays enter the subsequent analysis with a weight depending on
the gravitational lensing magnification at the back-projected origin of the photon packet.
}
\end{figure}

In this section we describe the general methodology for our calculations (see Figure~\ref{fig:sketch} for reference). 
In the first step, we use a ray tracing code which tracks photon packets from their origin 
in a lamppost corona or in the accretion disk forward in time until they reach a fiducial 
coordinate stationary observer at a distance of $r_{\rm obs}\,=$~10,000~$r_{\rm g}$.  
We denote the photon packets' wave vectors $k_{\rm r}^{\tilde{\mu}}$ in the reference frame of
a coordinate stationary receiver with tildes above the coordinate indices. Tracking photon packets forward in time 
allows us to account for multiple encounters of the photon packets 
with the accretion disk \cite[][and references therein]{2009ApJ...701.1175S}. 
The photons are collected at the fiducial observer and are used to generate a  {\it virtual polychromatic source} 
embedded in flat space time which mimics the original source embedded in the Kerr space time.
The virtual source is created by back-projecting the rays collected at $r_{\rm obs}$ 
onto a source plane at a distance of 10,000~$r_{\rm g}$ from the observer based on
the wavevector in the observer frame and assuming a flat space time geometry.
As the highly frequency shifted emission of interest here originates at distances of a few 
$r_{\rm g}$ from the black hole, the image seen by an observer at $r_{\rm obs}$ is (apart from the usual distance scaling) 
a good approximation of what  a much more distant observer would see.
In the second step we convolve the virtual source image with microlensing magnification
derived from the standard gravitational lensing theory \citep[see also][]{Popo:06,Jova:09,Chen:13,Nero:16}. 
\subsection{General Relativistic Ray Tracing Code}
\label{rt}
We use the general relativistic ray tracing code of \citet{2012ApJ...754..133K}, \citet{2016PhRvD..93d4020H}, and \citet{Behe:2016} to track photon packets from a lamppost corona to the observer. 
The code uses Boyer Lindquist (BL) coordinates $x^{\mu}=(ct,r,\theta,\phi)$ 
with $ct$ and $r$ in units of $r_{\rm g}$. The point-like lamppost corona is located at 
BL coordinates $r=h$, $\theta=\pi/18$ (close to the rotation axis of the black hole), 
and $\phi=0$ and emits photons isotropically in its rest frame. 
The  0-component of the photon packet's wave vector $k^{\mu}$ is initially normalized to $k^0=1$ to keep track 
of the net frequency shift between the emission and observation of the packet. 
The photon packets represent a power law distribution of initially unpolarized corona photons with 
differential spectral index $\Gamma$ (from $dN/dE\propto E^{-\Gamma}$). 
We assume a geometrically thin accretion disk extending from the innermost stable circular orbit 
$r_{\rm ISCO}$ to 100 $r_{\rm g}$ with the accretion disk matter orbiting the black hole on circular orbits.

The position and wave vector are evolved forward in time by integrating the geodesic equation:
\begin{equation}
\frac{d^2 x^\mu}{d \lambda ^{\prime 2}}= -\Gamma^\mu_{ \;\;\sigma \nu} \frac{d x^\sigma}{d \lambda ^\prime} \frac{d x^\nu}{d \lambda ^ \prime}.
\end{equation}
The $\Gamma^\mu_{ \;\;\sigma \nu}$'s are the Christoffel symbols and $\lambda^\prime$ is the affine parameter.  
The integration uses a fourth-order Runge-Kutta method making use of the conservation of the photons energy 
and angular momentum at infinity \citep[e.g.][Equations (7)-(12)]{Psal:12}.
The code keeps track of the photon's polarization by storing the polarization fraction and vector. 
The polarization vector is evolved forward in time with the parallel transport equation:
\begin{equation}
\frac{d f^{\mu}}{d\lambda'}\,=\,
-\Gamma^{\mu}_{\,\,\,\sigma\nu}
f^{\sigma}\frac{dx^{\nu}}{d\lambda'}.
\end{equation}  

Photon packets impinging on the accretion disk at $\theta=\pi/2$  are either absorbed or scatter, or 
prompt the emission of an Fe~K$\alpha$ photon.
We adopt a phenomenological parameterization for the relative probabilities of these three processes
with an absorption probability of $p_{\rm abs}$ per encounter,  and a ratio $R$ between the probabilities 
for scattering and for the production of a Fe~K$\alpha$ photon
\citep[see][for detailed treatments]{Matt:93,2005MNRAS.358..211R,2013ApJ...768..146G}. 
We use $p_{\rm abs}\,=\,0.9$ and $R=1$ if not mentioned otherwise. 

Scattering off the accretion disk is implemented by first transforming the photon packet's wave and polarization 
vectors from the BL coordinates into the reference frame of the accretion disk plasma \citep{2012ApJ...754..133K}. 
Subsequently, the photon packet scatters as described by the formalism of 
\citet{1960ratr.book.....C} for the reflection of polarized or unpolarized emission off an indefinitely 
deep electron atmosphere. After scattering, the photon packet's wave and polarization vectors 
are back-transformed into the BL coordinate frame.  

The emission of a mono-energetic Fe~K$\alpha$ photon is implemented in a similar fashion to the scattering.
After transforming the incomings photon packet's wave and polarization vectors into the reference 
frame of the accretion disk plasma, 
we use the $g$-factor between emission and absorption to weight the packet according to the 
assumed power law distribution of the 
lamppost corona emission. Subsequently, a mono-energetic Fe~K$\alpha$ 
photon packet is emitted with a limb brightening weight and an initial polarization given by Chandrasekhar's 
results for the emission from an indefinitely deep electron atmosphere \citep{1960ratr.book.....C}. 
In the final step,
the photon packet's wave and polarization vectors are backtransformed into the global BL frame.

As BL coordinates exhibit a coordinate singularity at the event horizon, they cannot be used to transport photon packets 
across the event horizon. We stop tracking packets when their radial coordinate gets closer than 
1.02 times the $r$-coordinate of the event horizon assuming that they will not escape the black hole. 

For the rest of the paper, we focus our attention on the Fe~K$\alpha$ packets, 
assuming that we will compare the energy spectrum of the simulated Fe~K$\alpha$ photon packets
with observed energy spectra after subtracting the continuum contribution.
\subsection{The Effect of Gravitational Lensing}
\label{micro}
As mentioned above, we discuss the gravitational lensing of the Fe~K$\alpha$ emission for the specific case of
RX~J1131$-$1231. The lens geometry depends on the angular diameter distance of the source 
$D_{\rm S}$,  the lens $D_{\rm L}$, and between the lens and the source $D_{\rm LS}$. 
For an observer at redshift $z_1$ the angular diameter of a source at redshift $z_2$ is given 
by \citep[see][]{Peeb:93,Hogg:00}:
\begin{equation}
D(z_1,z_2)\,=\,\frac{1+z_1}{1+z_2}\,\frac{c}{H(z_1)} \, \int_{z_1}^{z_2} \frac{dz}{E(z)}
\end{equation}
with
\begin{equation}
E(z)\,=\sqrt{\Omega_{\rm m}(1+z)^3 +\Omega_{\rm \Lambda}}
\end{equation}
and 
\begin{equation}
H(z)\,=\,H_0\,E(z).
\end{equation}
The equations assume a flat universe, and we use the 2015 Planck cosmological parameters, 
i.e.\  a Hubble constant $H_0\,=$~67.74 km s$^{-1}$ Mpc$^{-1}$, and the matter and 
dark energy densities of $\Omega_{\rm m}\,=$~30.89\% and $\Omega_{\Lambda}\,=$~ 69.11\% in units of the
critical density, respectively \citep{Plan:15}.
For  RX~J1131$-$1231, we obtain 
$D_{\rm L}\,=$~0.94~Gpc,  $D_{\rm LS}\,=$ 0.83~Gpc, and $D_{\rm S}\,=$~1.48~Gpc.

Microlensing by deflectors of average mass $<\!\!M\!\!>$ leads to typical light deflections on the order of:   
\begin{equation}
\alpha_0\,=\,\sqrt{\frac{4\,G\,<\!\!M\!\!>}{c^2} \frac{D_{\rm LS}}{D_{\rm L}D_{\rm S}}}.
\end{equation}
The corresponding distance in the source plane, the Einstein radius, is given by: 
\begin{equation}
\zeta_0\,=\,  \alpha_0\,D_{\rm S}.
\label{einsteinRadius}
\end{equation}
With $<\!\!\!M\!\!\!>$~$=\,0.25\, M_{\odot}$ we get $\zeta_0\,=$~2.44$\times 10^{16}$~cm
which equals 1,654~$r_{\rm g}$ assuming a black hole mass of $10^8\,M_{\odot}$.

As mentioned above we use a simple parameterization of the magnification close to caustic folds in this first paper.
The approach has the advantage of avoiding a systematic exploration of the  microlensing parameter space. 
The magnification close to a caustic fold can be derived by expanding the Fermat potential into a Taylor 
series and retaining only leading order terms \citep{1992grle.book.....S}. 
Close to the caustic fold, the magnification $\mu$ relative to the magnification 
outside the caustic $\mu_0$ is given by the expression: 
\begin{equation}
\mu/\mu_0\,=\,1+\frac{K}{\sqrt{y_{\perp}}}H(y_{\perp})
\label{EQamp}
\end{equation}
with $y_{\perp}$ being (up to the sign) the distance from the fold located at $y_{\perp}\,=\,0$, 
the caustic amplification factor $K$, and the Heaviside step function $H(y_{\perp})\,=\,0$ 
for $y_{\perp}<0$ and $H(y_{\perp})\,=\,1$  for $y_{\perp}\ge 0$.
In the following, we call the side with $y_{\perp}>0$ and $\mu/\mu_0>1$ the positive side of the caustic fold.
In the general case, $K$ can be derived from partial derivatives of the Fermat potential. 
In the particular case of lensing through a random field of stars, $K$ is given by
\begin{equation}
K\,=\,\beta \sqrt{\zeta_0}
\label{eqn:K}
\end{equation}
with $\beta$ being a constant of order unity \citep[][]{1993A&A...268..501W,2002ApJ...568..509C}.
Remarkably, using Equation (\ref{EQamp}) with $y_{\perp}$ in units of $r_{\rm g}$, the magnification 
depends only on one adjustable parameter. For RX~J1131$-$1231 and a typical $\beta$-value of 0.5,
we can relate the width of the caustic fold (perpendicular to the fold) to the size of the accretion disk 
(with an inner diameter of a few $r_{\rm g}$) by noting that the magnification exceeds $\mu_0$ 
by more than one order of magnitude over a distance of $\approx4\,r_{\rm g}$ perpendicular to the caustic. 
As mentioned above, the Einstein radius of 0.25 solar mass stars projected onto the source plane is 
1,654 $r_{\rm g}$, and the straight fold approximation holds. The folds typically extend over thousands 
of $r_{\rm g}$ \citep[see for example the inverse ray shooting results of][]{Dai:10}.%
\section{Results}
\label{results}
We ran the simulations for black hole spin values $a$ from 0 to 0.9 in steps of 0.1 
plus additional simulations for $a\,=$~0.95, 0.98, and 0.998. Furthermore, we simulated 
lamppost heights $h/r_{\rm g}$ of 5,10, 20, and 40 (with a few additional simulations at 
$h/r_{\rm g}\,=50, 100$). The data were analyzed 
in 5$^{\circ}$ wide inclination bins and for $\beta$-values of 0.125, 0.25, 0.5, 1, and 2. 
For each simulated black hole spin, we simulate 
caustic crossings for crossing angles $\theta_{\rm c}$ (the angle between the normal of the caustic fold 
and the black hole spin axis) between 0 and $360^{\circ}$ in $20^{\circ}$ steps and for 
-30~$r_{\rm g}$ to +30~$r_{\rm g}$ offsets of the caustic from the center of the black hole
in steps of 0.1~$r_{\rm g}$. An angle of $\theta_{\rm c}\,=\,0$ corresponds to a caustic fold 
perpendicular to the black hole spin axis with the positive side of the caustic pointing 
in the $\theta\,=\,0$ direction. If not mentioned otherwise,  we show the results for the 
fiducial values $a\,=\,0.5$, $i\,=\,82.5^{\circ}$, $h\,=\,5\,r_{\rm g}$, 
and $\beta\,=\,0.5$.

\begin{figure}
\plotone{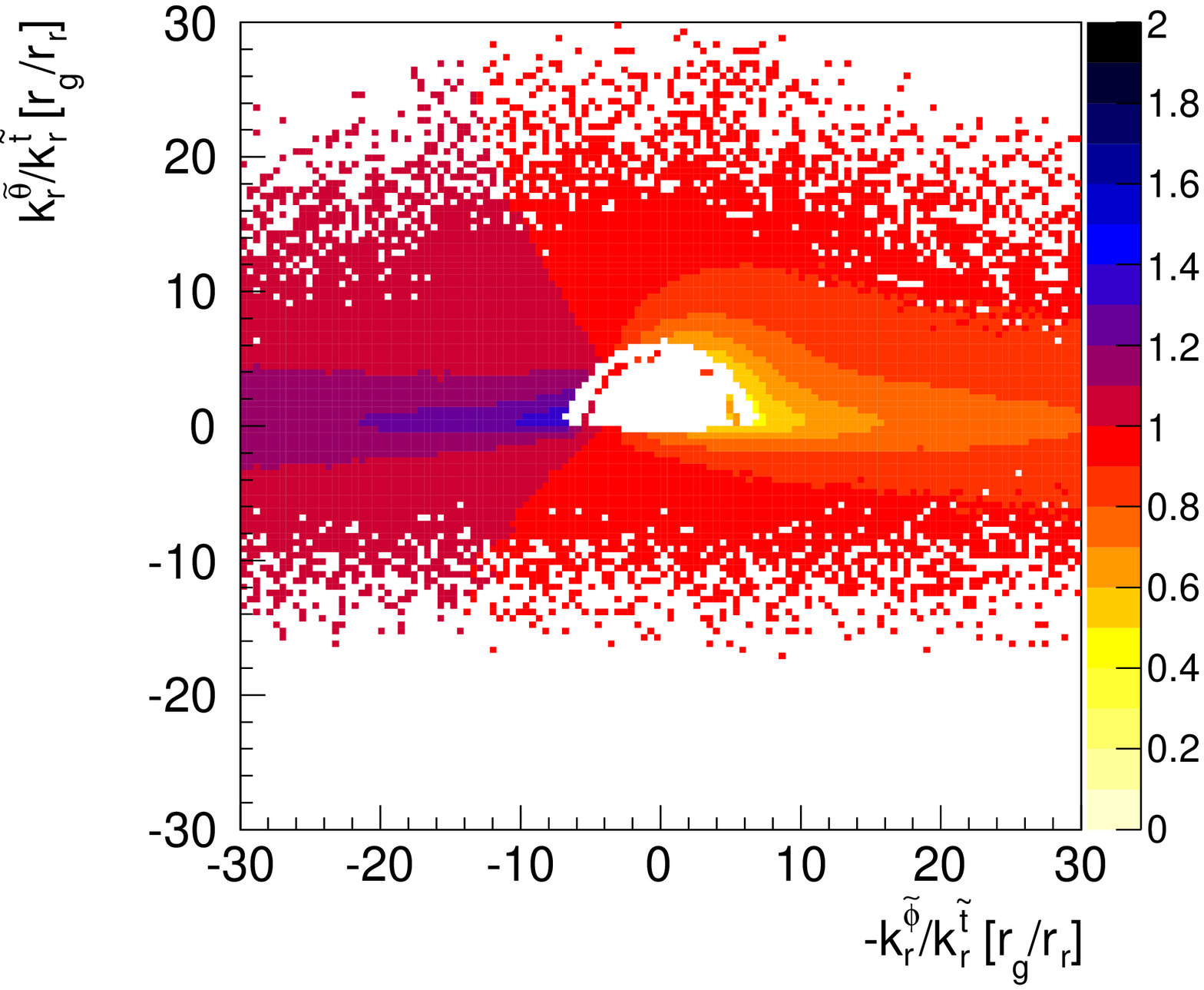}
\plotone{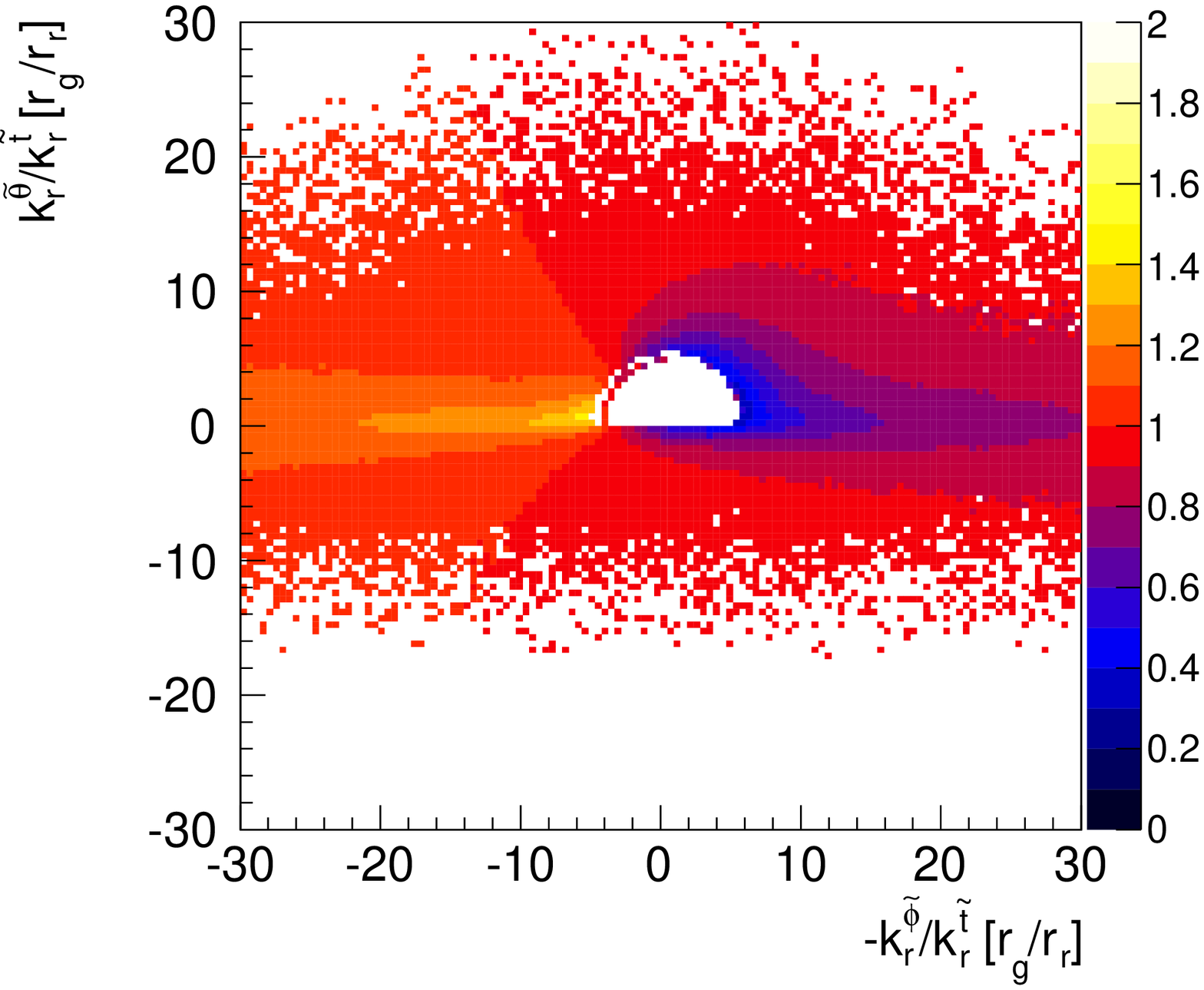}
\plotone{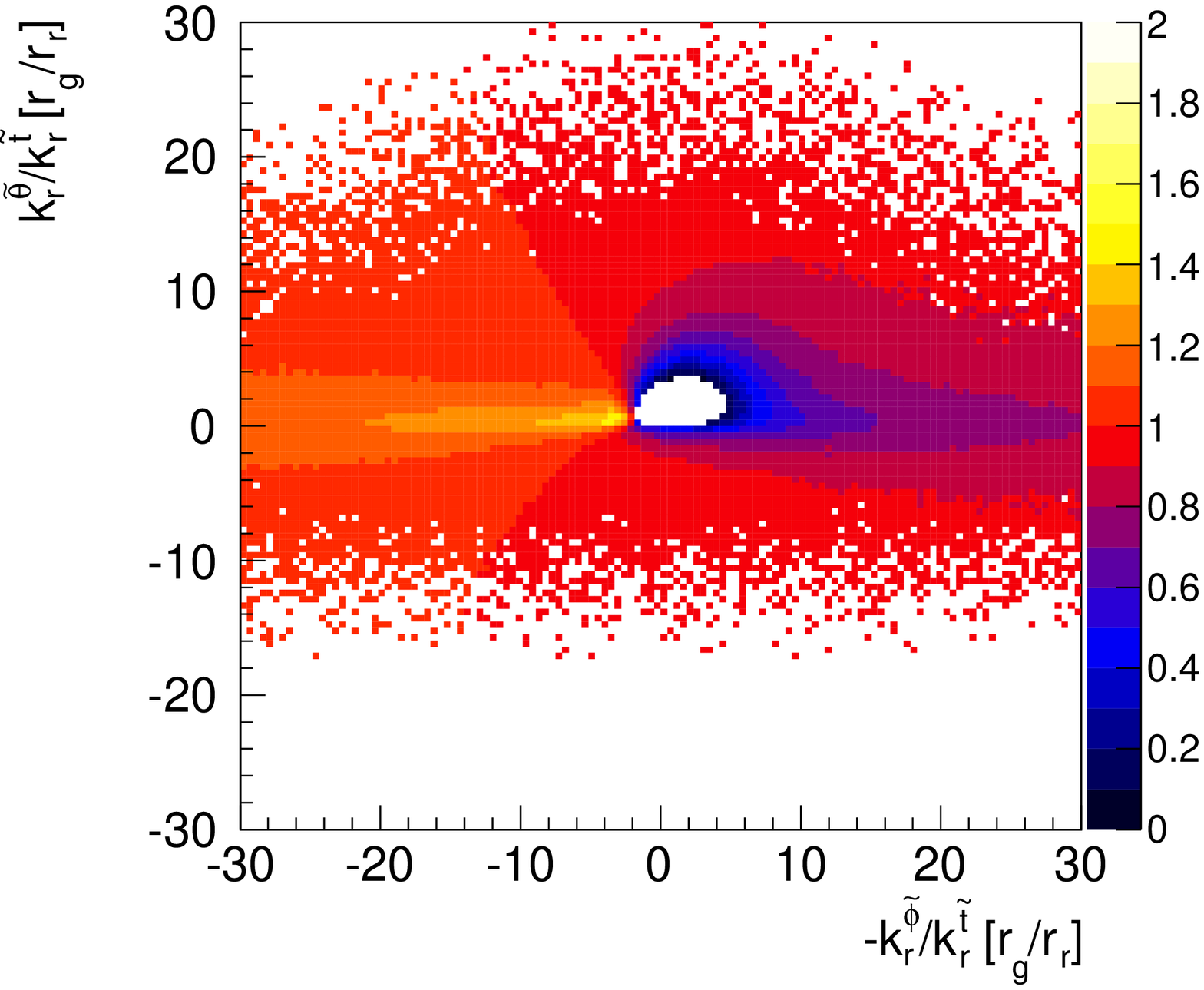}
\caption{\label{fig:gfactors} Maps of the $g$-factors of the Fe~K$\alpha$ emission originating 
in the accretion disks of black holes with spins $a\,=\,0$ (top), $a\,=\,0.5$ (center), and $a\,=\,0.998$ (bottom) 
All panels: $h\,=\,5\,r_{\rm g}$, $i\,=\,82.5^{\circ}$, $\beta\,=\,0.5$. 
Here and in the following, $k_{\rm r}^{\tilde{\mu}}$ denotes the photon packets' received wavevector (subscript $r$) 
in the reference frame of a coordinate stationary observer (marked with a tilde above the coordinate index).      
}
\end{figure}

Figure~\ref{fig:gfactors} shows 2-D maps of the $g$-factors of the 
Fe~K$\alpha$ emission for BHs with $a\,=\,0,\,0.5,\,0.998$ ($i\,=\,82.5^{\circ}$). 
The maps clearly show the effects of the relativistic motion of the accretion disk plasma 
and the curved spacetime. The Doppler boosting of the plasma moving towards (away from) 
the observer gives $g$-factor larger (smaller) than unity. Close to the event horizon, 
gravity always wins over the Doppler effect leading to $g\ll 1$.
For the lower black hole spins, the photon ring between the black hole and the accretion disk can be recognized 
resulting from photons orbiting the black hole for $n\times \pi$ turns ($n\ge1$) before escaping the black hole's
gravitational attraction.

\begin{figure}
\plotone{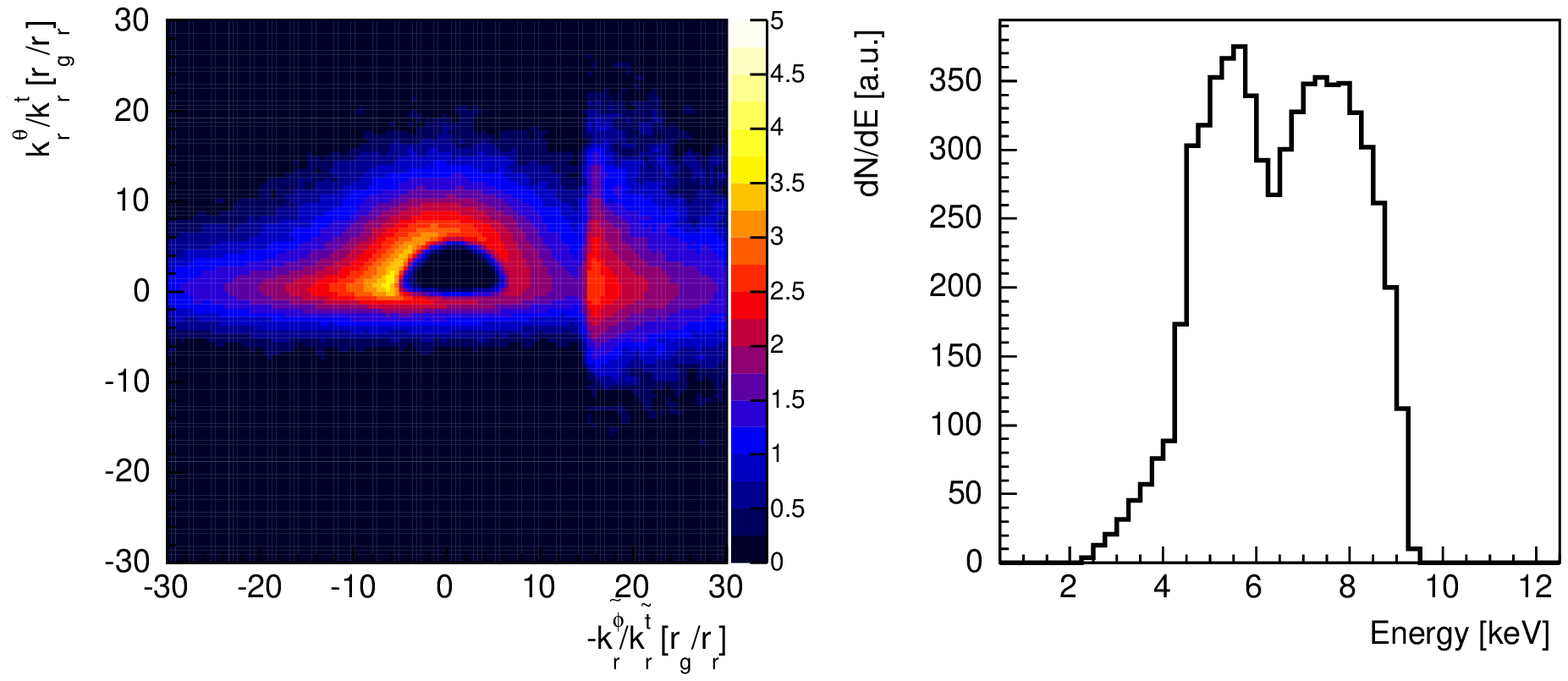}\plotone{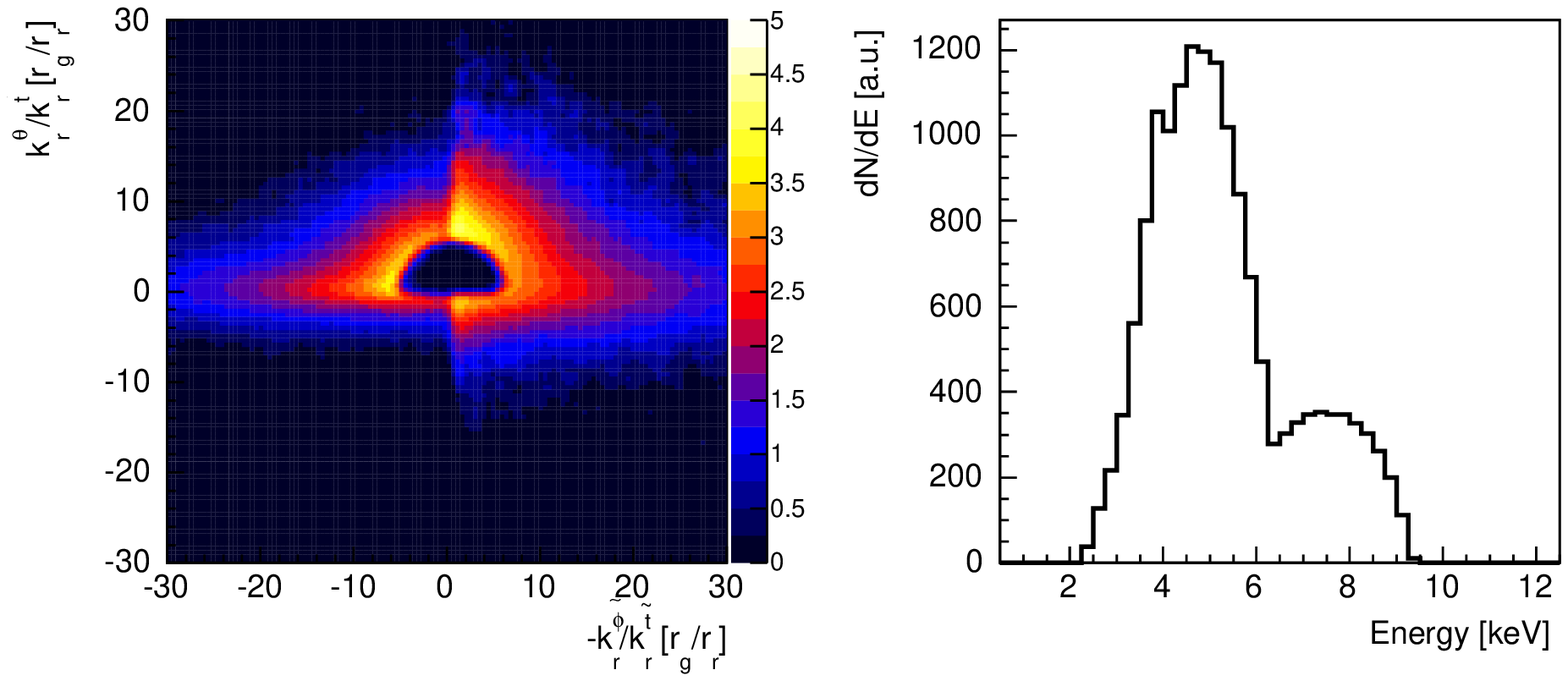}
\plotone{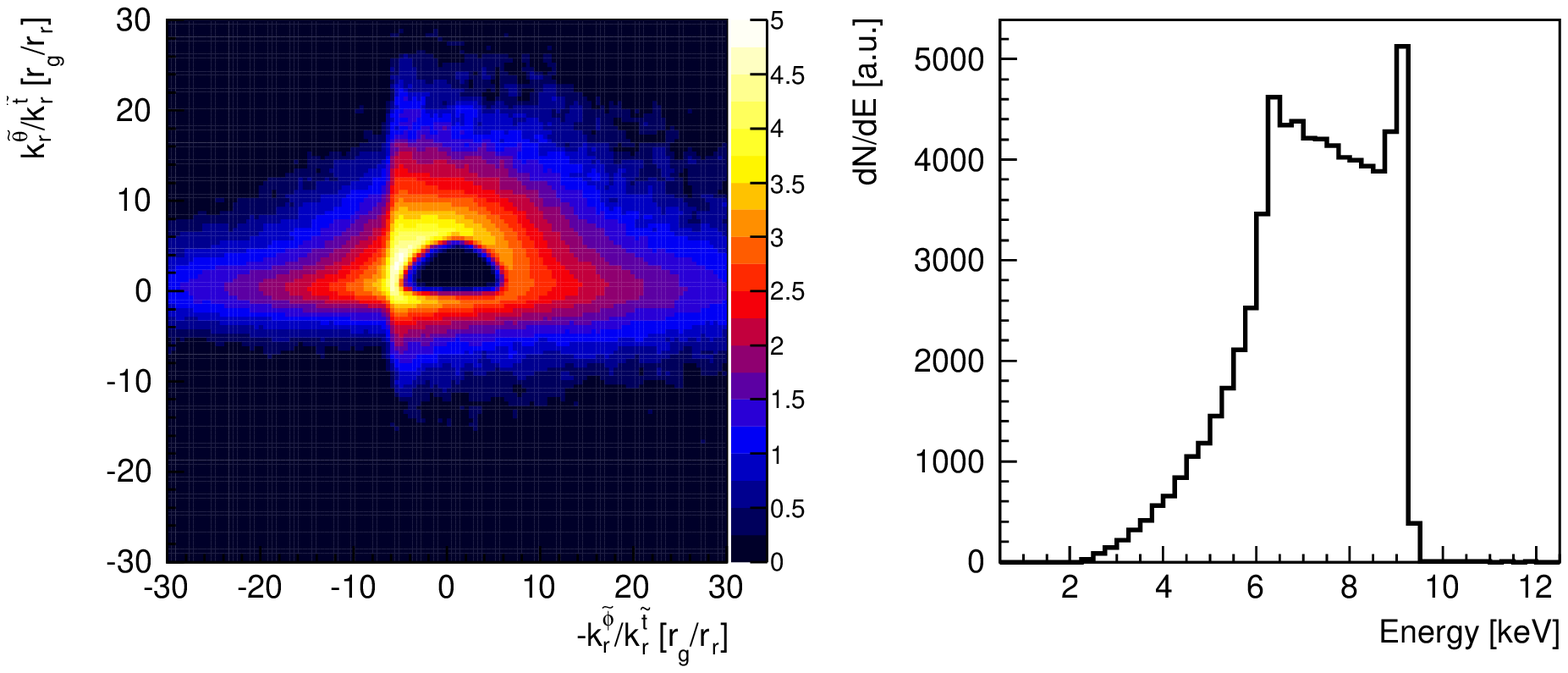}\plotone{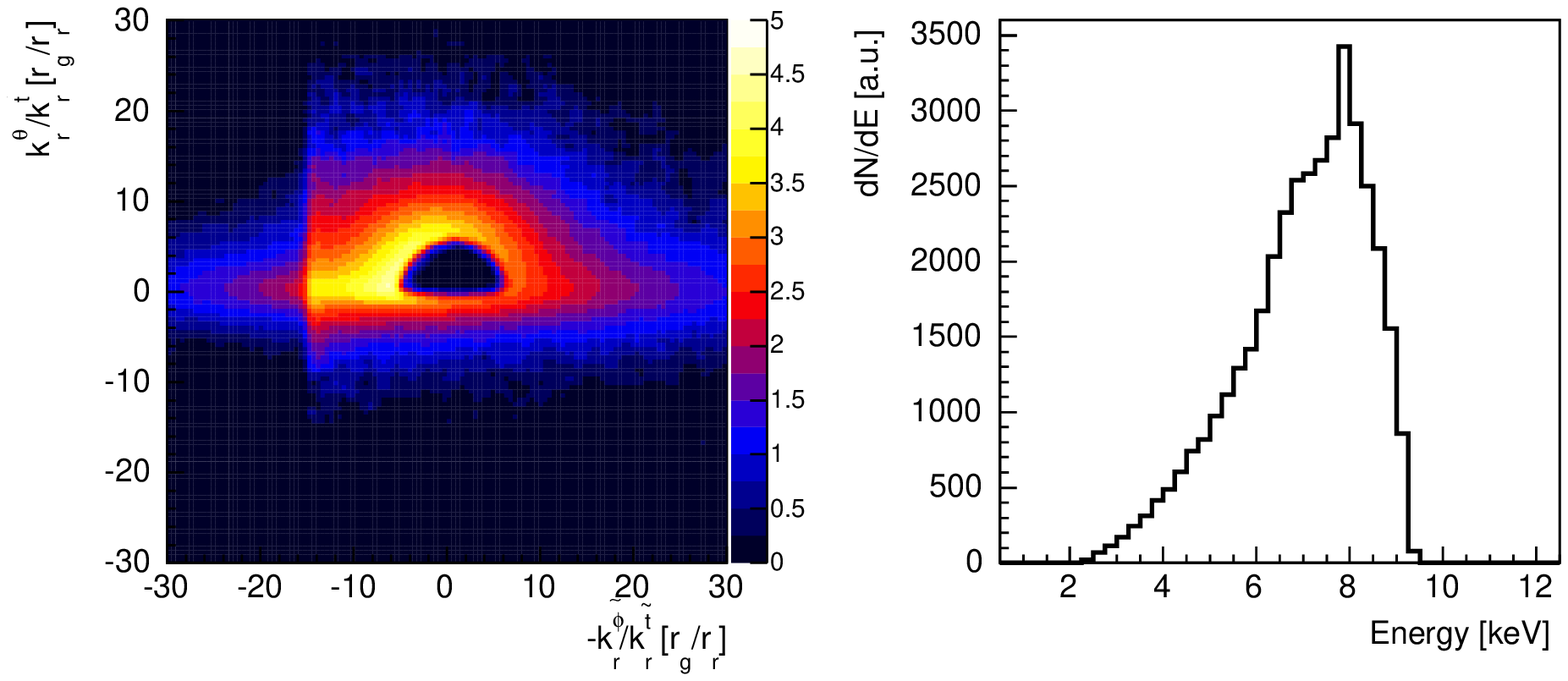}
\caption{\label{fig:m01} Four snapshots of a caustic with caustic crossing angle $\theta_{\rm c}\,=\,90^{\circ}$ moving from the
right to the left. The left panels show the magnification weighted surface brightness with a logarithmic color scale.
The right panels show the corresponding energy spectra of the Fe~K$\alpha$ emission.
All panels: $a\,=\,0.5$, $i\,=\,82.5^{\circ}$, $\beta\,=\,0.5$.   
}
\end{figure}

\begin{figure}
\plotone{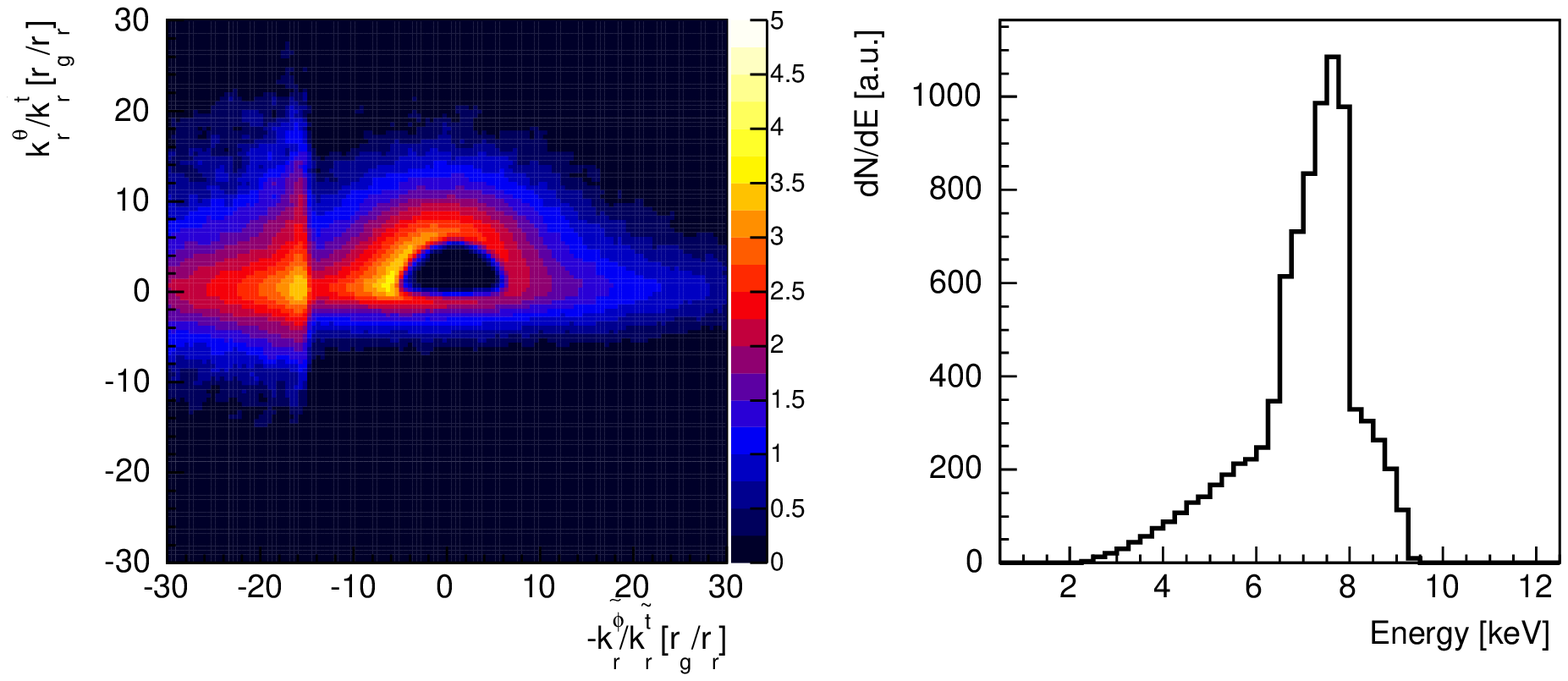}\plotone{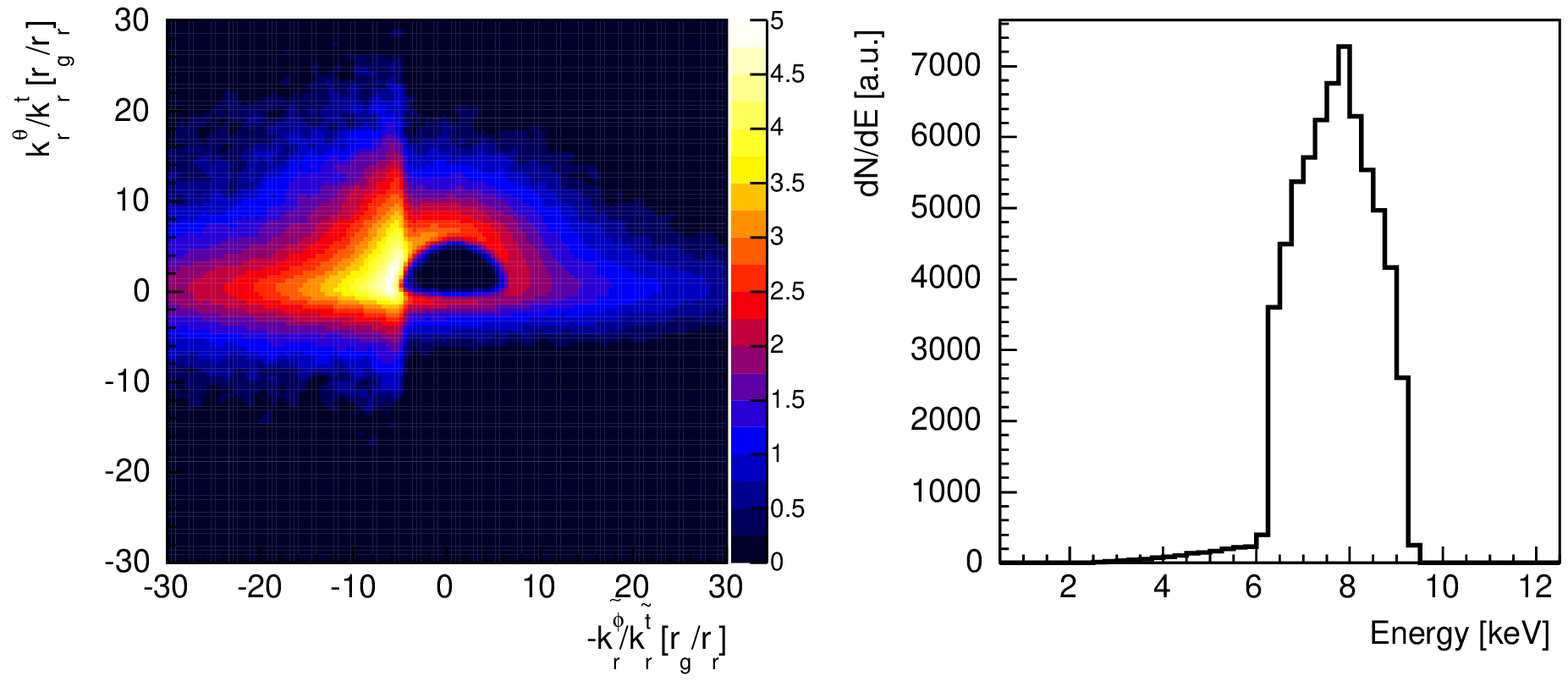}
\plotone{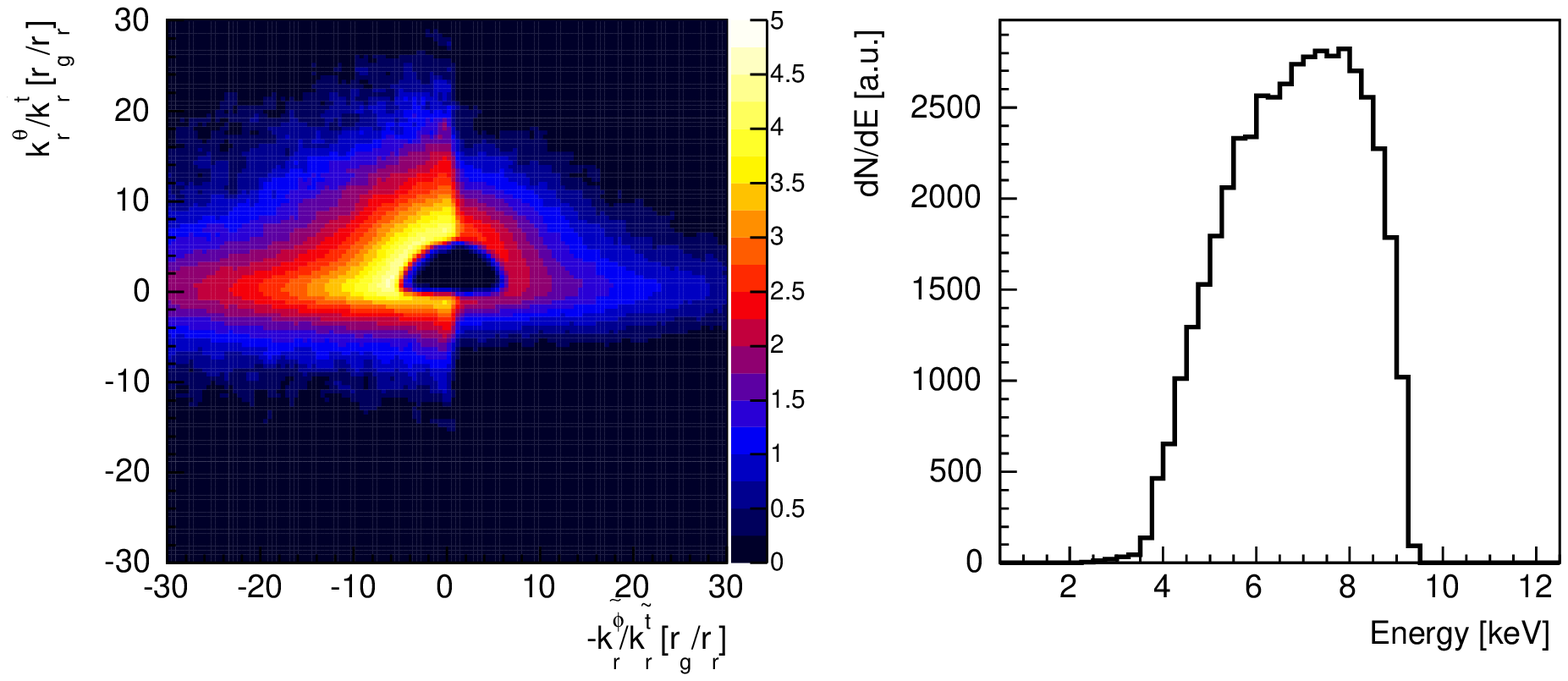}\plotone{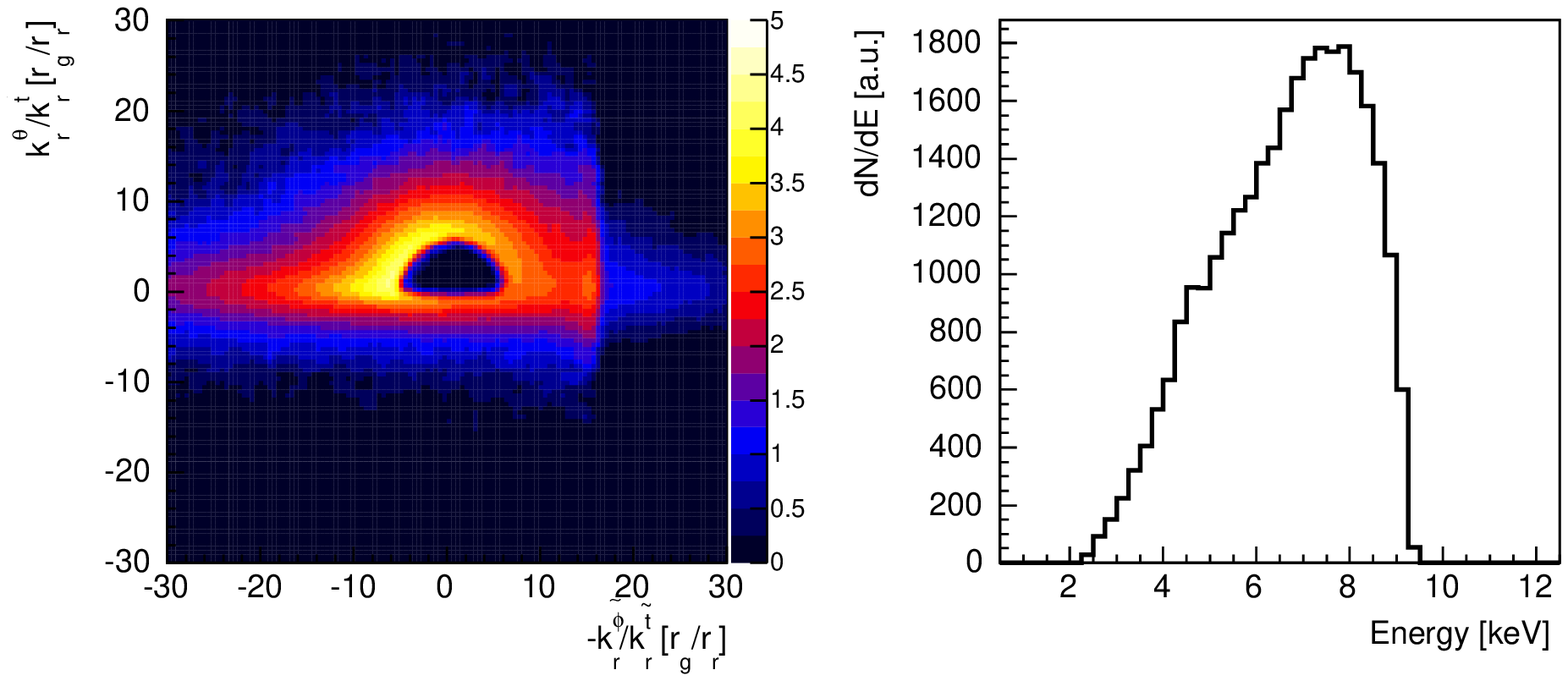}
\caption{\label{fig:m02} Same as Figure~\ref{fig:m01} but for the opposite orientation of the caustic, i.e. for  
$\theta_{\rm c}\,=\,270^{\circ}$. Note that the amplification patterns are independent of the direction of the movement of the caustic.
The magnification patterns of Figures~\ref{fig:m01} and \ref{fig:m02} could both run from the 
left to the right or vice versa.
}
\end{figure}

Figures~\ref{fig:m01} and \ref{fig:m02} shows the 2-D surface brightness distributions (left panels) 
and the surface brightness and microlensing magnification weighted energy spectra 
of the Fe~K$\alpha$ emission (right panels) for two opposite caustic orientations 
$\theta_{\rm c}\,=\,90^{\circ}$ and 
$\theta_{\rm c}\,=\,270^{\circ}$, respectively.  
Note that an observer with an excellent angular resolution telescope would see different surface brightness 
distributions owing to the image distortions caused by the gravitational lens. 
The effect of the caustic can clearly be recognized by a jump of the surface brightness 
when the magnification increases from $\mu/\mu_0\,=\,1$ to $\mu/\mu_0\gg1$  
followed by the gradual return to $\mu/\mu_0\,\approx\, 1$.  
Both caustic orientations shown in Figures~\ref{fig:m01} and \ref{fig:m02} produce complex Fe~K$\alpha$ 
energy spectra when the caustic fold gets close to the brightest part of the accretion disk approaching 
the observer with a velocity close to the speed of light. Of particular interest is the third panel of 
Figure~\ref{fig:m01} where the amplification of the brightest part of the accretion disk leads to a 
high-intensity double-peaked energy spectrum which would be readily detectable. 
Comparing the energy spectra for the two caustic orientations shown in Figures~\ref{fig:m01} and \ref{fig:m02}
one can recognize that the former leads more often to double-peaked Fe~K$\alpha$ energy spectra than the latter.
Double peaked energy spectra result naturally when the caustic fold primarily magnifies the emission from the 
receding part of the accretion disk. The low-energy peak originates from the magnification of the redshifted 
dimmer emission,  and the bright strongly Doppler-boosted part of the accretion disk dominates 
the high-energy peak. In contrast, the caustic orientation of Figure~\ref{fig:m02} leads to fewer 
double-peaked energy spectra as the amplification of the brightest part of the disk magnifies 
the spectral peak which dominates the energy spectrum anyhow, even without caustic magnification.  
It is interesting to note that the microlensing amplification of the approaching and receding parts of the accretion
disk emission has been invoked in the context of explaining the fine structure in the light curves of 
high-amplification events from gravitationally lensed quasars \citep[see][]{Abol:12,Medi:15}.

For each simulated data set we analyze the energy spectra for all simulated caustic crossing angles 
and offsets with an algorithm identifying the most prominent peaks in the photon number energy spectra ($dN/dE$)
and fitting them with Gaussian distributions. The algorithm first finds the highest peak and subsequently searches 
for an additional peak rising $\ge10$\% of the amplitude of the brightest peak above the valley between 
the two peaks. We call energy spectra with one peak singles (with a spectral peak at energy $E_1$) 
and energy spectra with two peaks doubles
(with spectral peaks at $E_1$ and $E_2$, $E_2<E_1$). The following figures show the distributions 
of the fitted parameters.   

\begin{figure}
\plotone{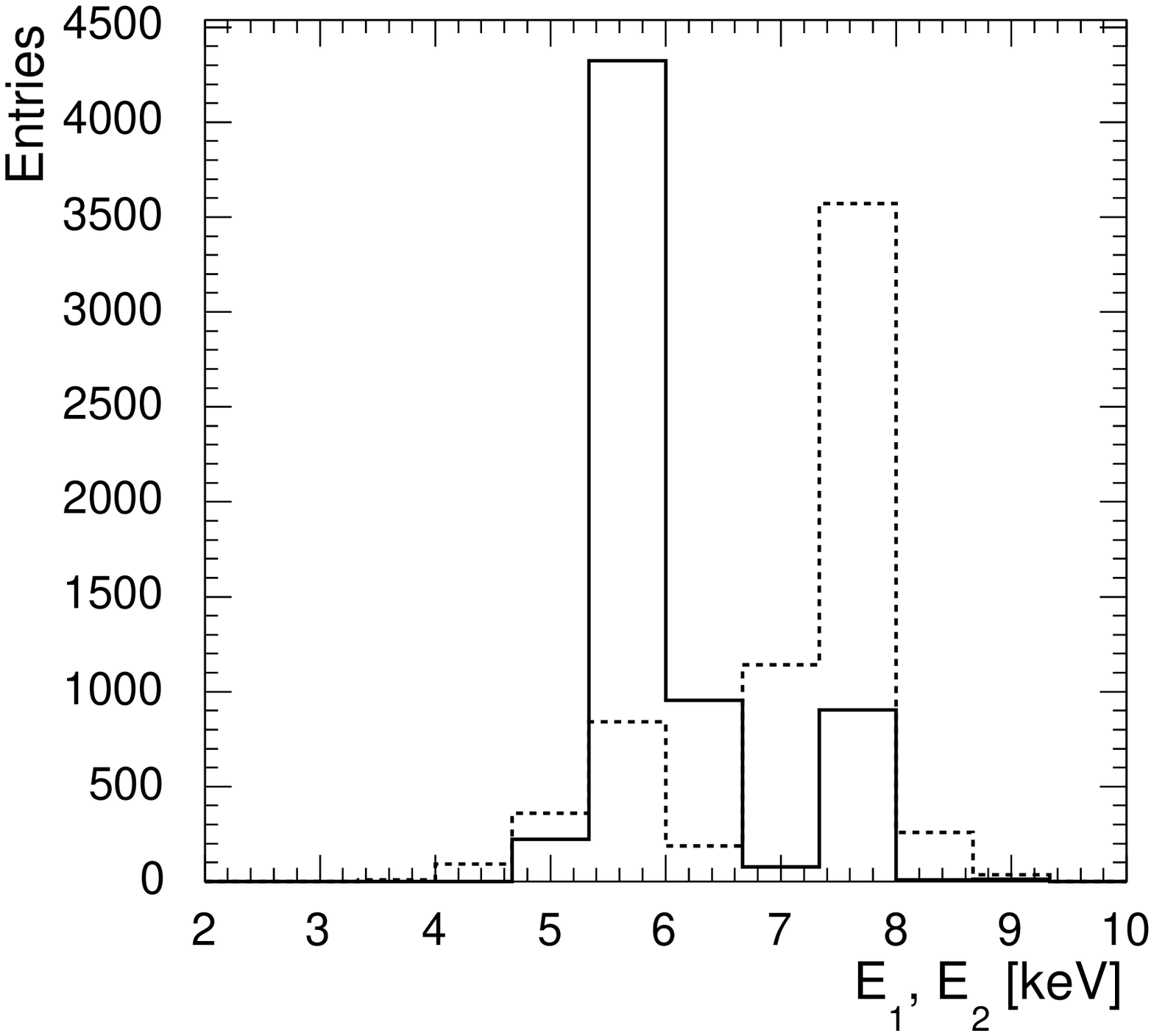}
\plotone{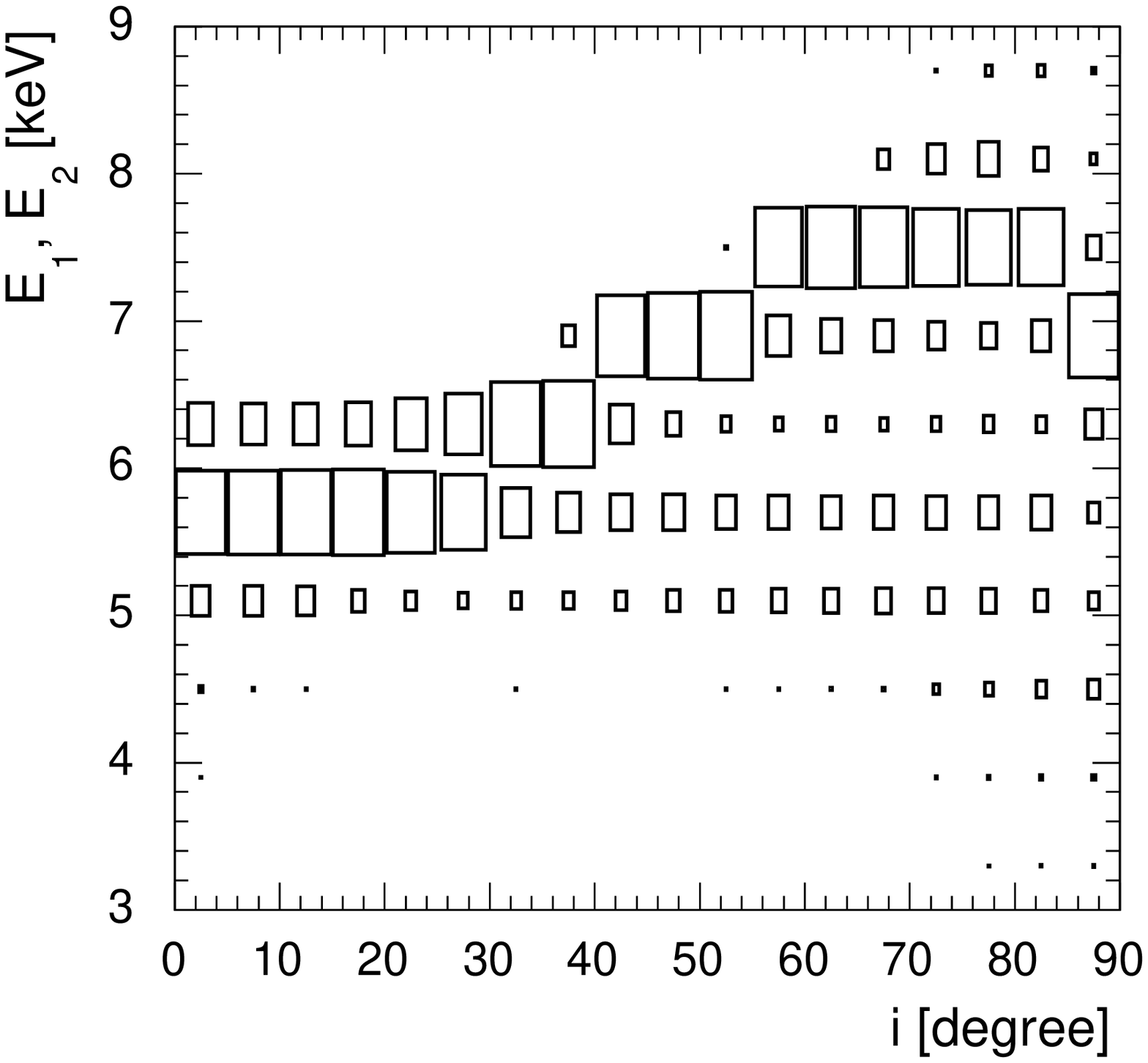}
\caption{\label{fig:inclination} The upper panel shows the spectral peak energies for inclinations
of $12.5^{\circ}$ (solid line) and $82.5^{\circ}$ (dashed line) 
for all simulated caustic crossing angles and caustic offsets. 
The lower panel shows the spectral peak energies as a function of inclination $i$ 
(the box size is proportional to the number of simulated events).
Both panels: $a\,=\,0.5$, $h\,=\,5\,r_{\rm g}$,
$\beta\,=\,0.5$.}
\end{figure}

The upper panel of Figure~\ref{fig:inclination} shows the distribution of the peak energies (singles and doubles) 
for a black hole with spin $a\,=\,0.5$ for two inclinations: $i\,=\,12.5^{\circ}$ and $i\,=\,82.5^{\circ}$.
For the former, the distribution is narrowly peaked with a minimum and maximum energy of 4.7 and 8~keV, respectively.  
For the latter, the distribution is shifted to higher energies and is much broader with a minimum and maximum energy 
of 4 and 9.0~keV, respectively. The lower panel of Figure~\ref{fig:inclination} presents the peak distribution 
as a function of inclination. The gradual shift of the peak energies towards higher energies and the broadening 
of the distribution can clearly be recognized.  

\begin{figure}
\plotone{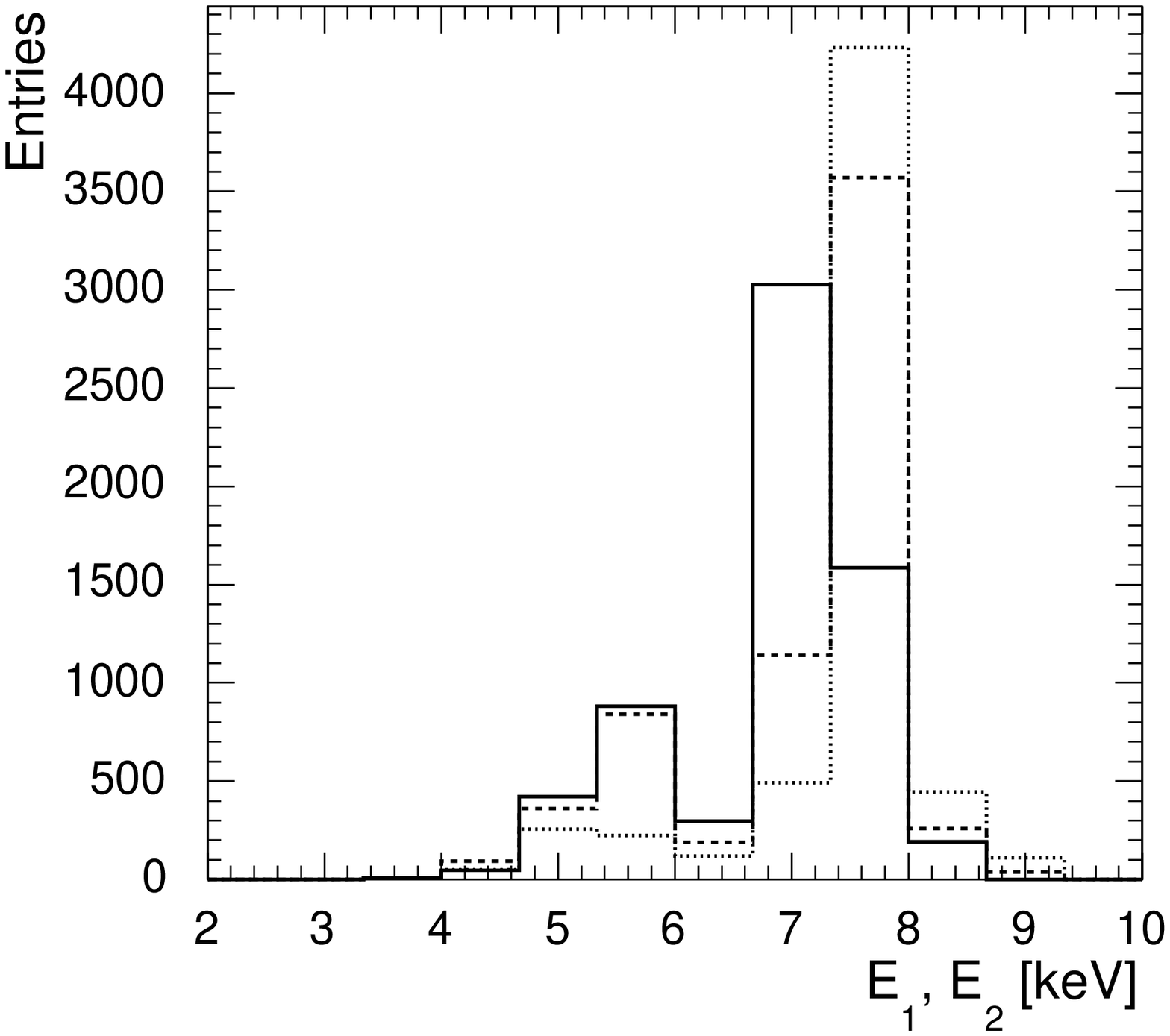}
\plotone{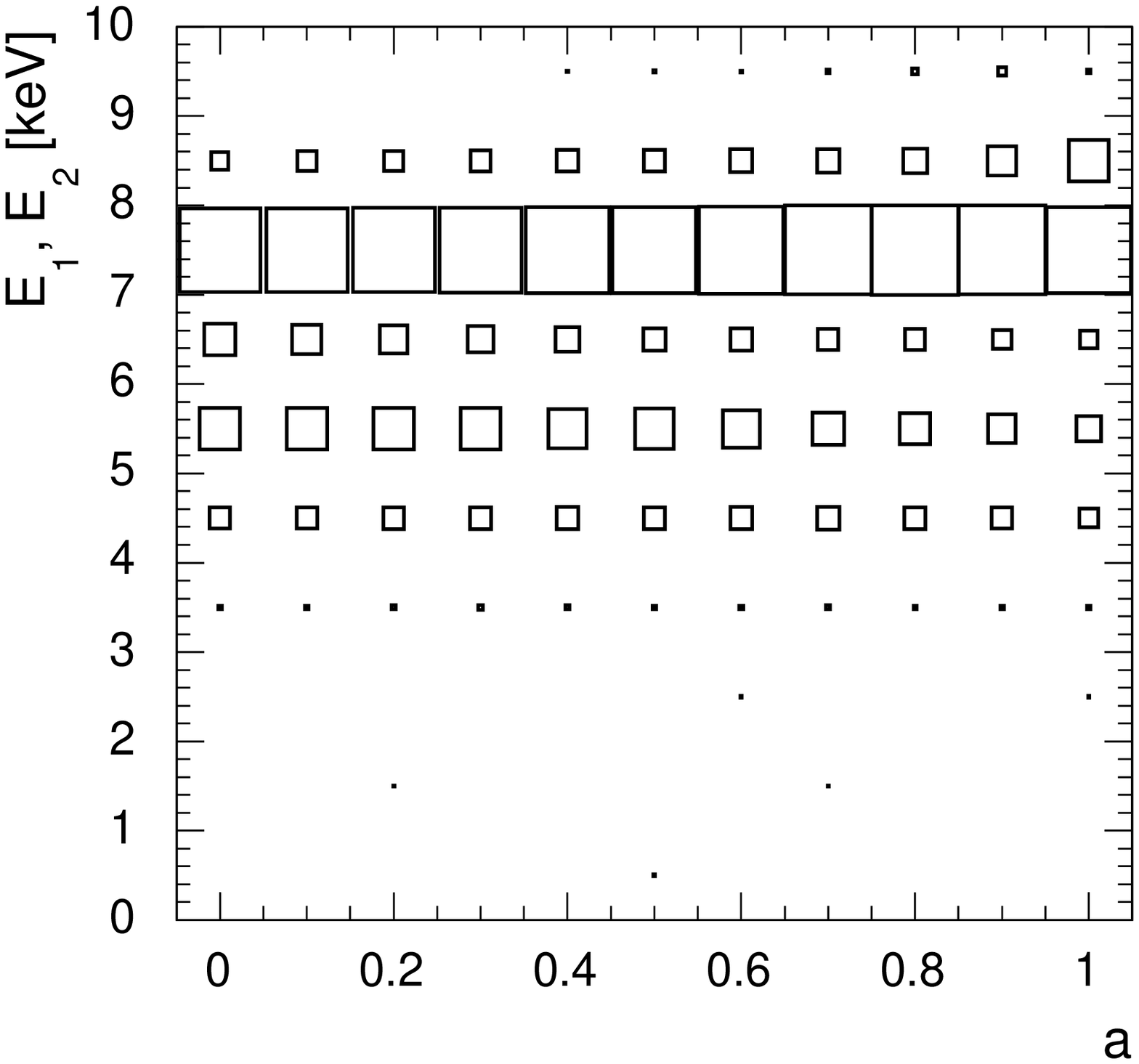}
\caption{\label{fig:singles} The upper panel shows the spectral peak energies 
for black hole spins $a\,=\,0$ (solid line), $a\,=\,0.5$ (dashed line), $a\,=\,0.998$ (dotted line).
The lower panel shows the spectral peak energies as a function of $a$.
Both panels: $i\,=\,82.5^{\circ}$, $h\,=\,5\,r_{\rm g}$,
$\beta\,=\,0.5$.}        \vspace*{-2ex}

\end{figure}

Figure~\ref{fig:singles} presents the dependence of the peak energies on the black hole spin for the 80$^{\circ}$ 
to 85$^{\circ}$ inclination bin.  Although the peak energies shift towards higher energies with increasing spin, 
the effect is rather subtle. 
Figures~\ref{fig:inclination} and \ref{fig:singles} confirm the conclusions in C17 that 
the range of detected spectral peaks, a rather robust observational result, can be used to constrain the inclination 
of the accretion disk. Constraining the black hole spin requires to compare the full distributions of the
detected spectral peak energies and thus a more careful match of simulations and data. 

Next we analyzed all energy spectra with two spectral peaks, and checked if the energy separations $\Delta E\,=\,E_2-E_1$, 
the ratio of the line fluxes $\rho_n\,=\,n_2/n_1$ (from fitting the peaks with Gaussians), the widths $w_1$ and $w_2$,  or the ratio of the widths 
$\rho_{\rm w}\,=\,w_2/w_1$ depend on the black hole spin $a$ and can thus be used to constrain it observationally.

\begin{figure}
\plotone{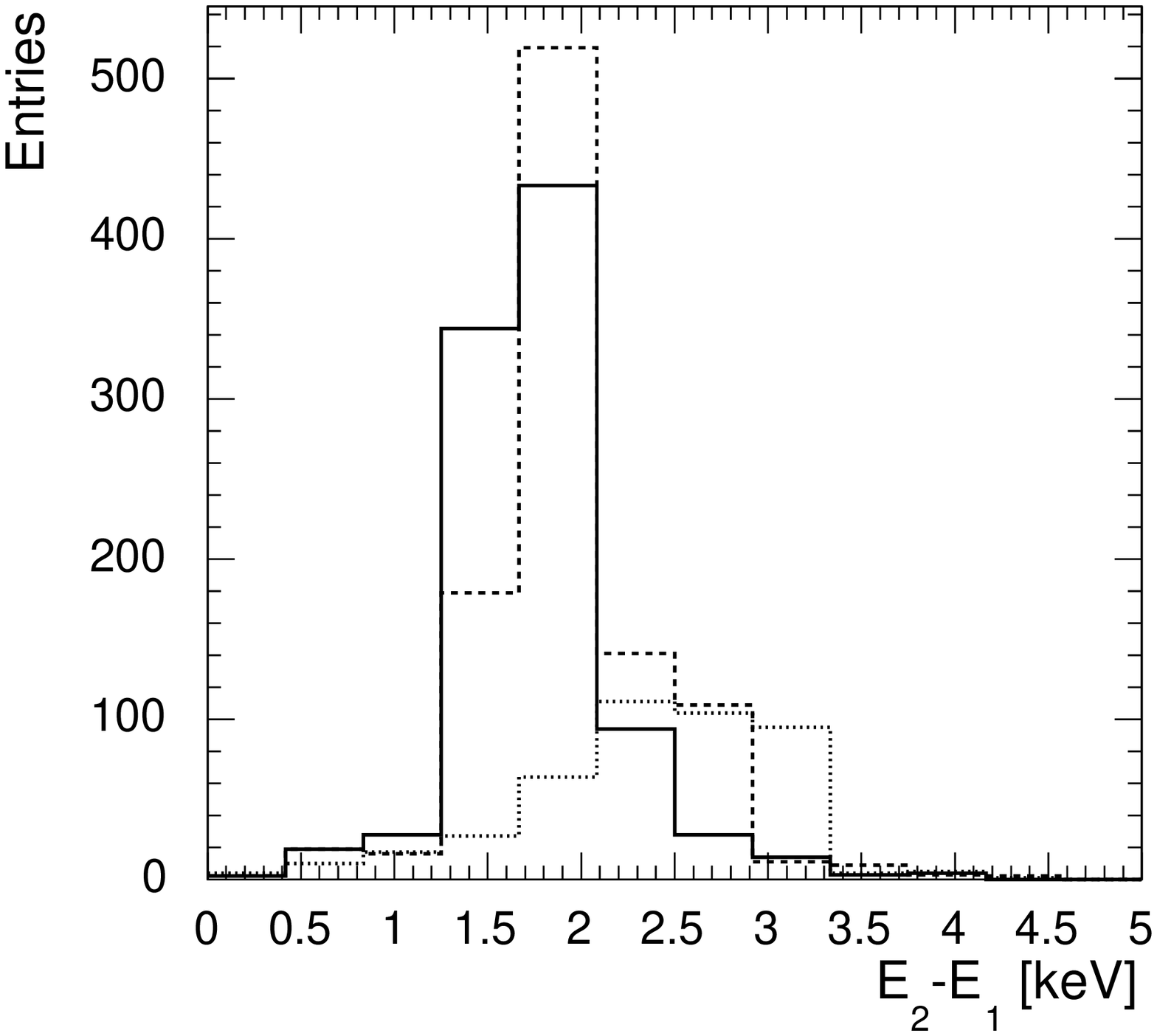}
\plotone{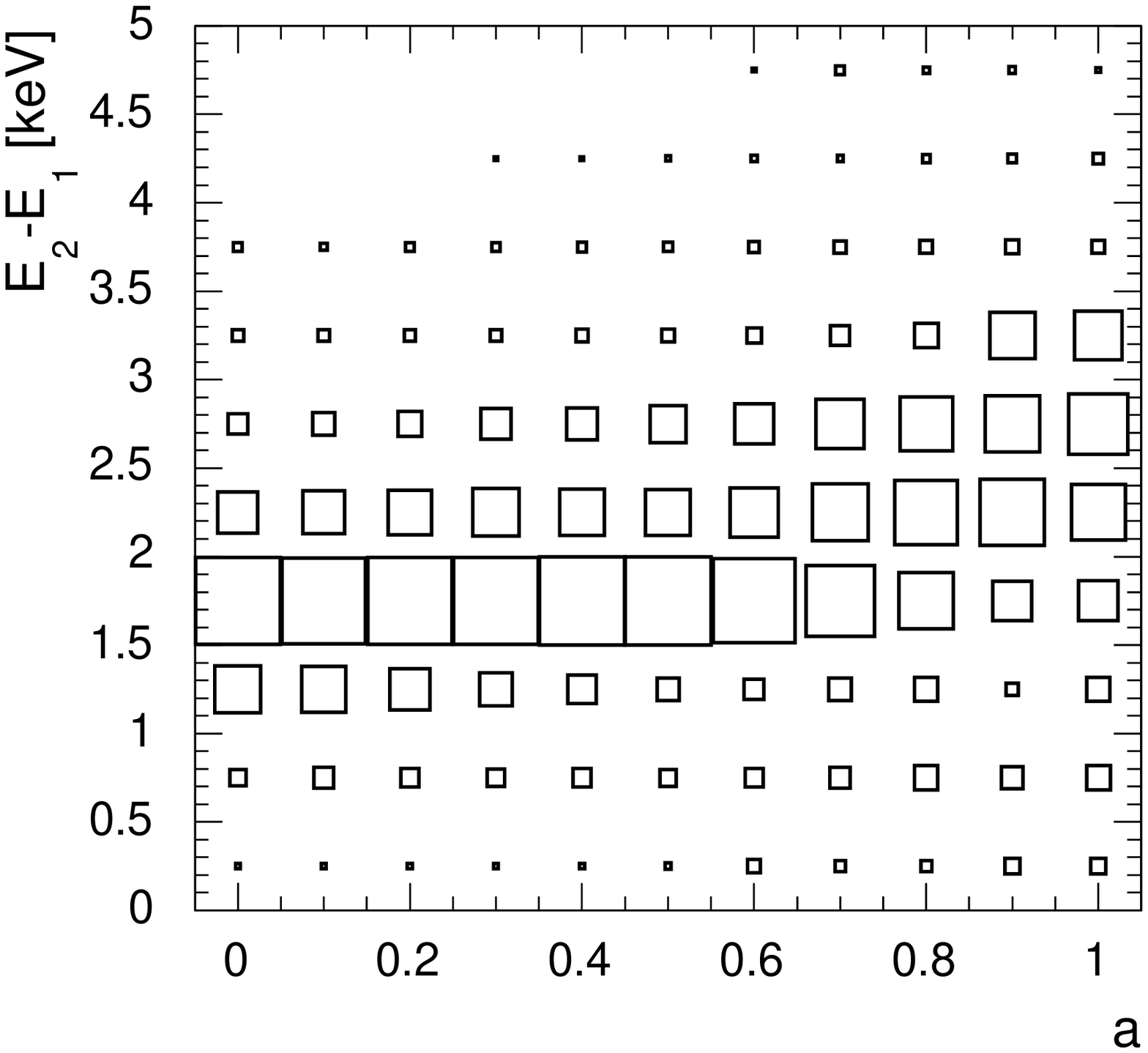}
\caption{\label{fig:delta} The upper panel shows the separation of the line centroids of all doubles
for black hole spins $a\,=\,0$ (solid line), $a\,=\,0.5$ (dashed line), $a\,=\,0.998$ (dotted line).
The lower panel shows the separations as a function of $a$. Both panels: $i\,=\,82.5^{\circ}$, 
$h\,=\,5\,r_{\rm g}$, $\beta\,=\,0.5$.}
\end{figure}

We find that the energy separations $\Delta E$ show the strongest dependence on $a$ with 
$\Delta E$, with $\Delta E$ increasing for higher $a$-values (Figure~\ref{fig:delta}). 
We presented the observed distribution of the $\Delta E$-values for RX~J1131$-$1231 in C17, 
noting that the peak of the distribution at 3.5$\pm$0.2 keV suggests a high value of the spin $a \simgt 0.8$.

The other quantities exhibit correlations with $a$ but only when accounting for correlations with other parameters. 
The maximum likelihood method presented in the next sub-section is able to 
take advantage of such subtle correlations.

\begin{figure}
\plotone{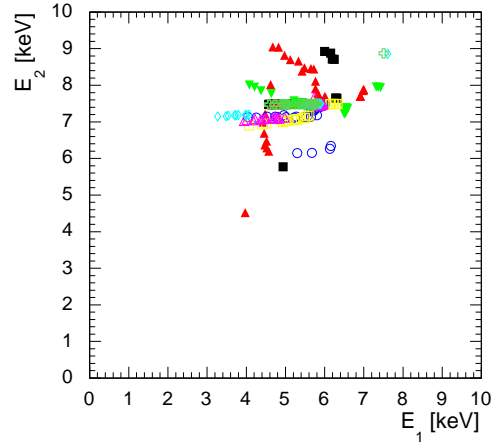}
\caption{\label{fig:correlation} Correlation of the peak energies of all doubles for a black hole with spin $a\,=\,0.5$ seen at
an inclination of $i\,=\,82.5^{\circ}$ ($h\,=\,5\,r_{\rm g}$, $\beta\,=\,0.5$). 
The different marker types (and colors in the online version) highlight
the values from different caustic crossing angles.}
\end{figure}

Figure~\ref{fig:correlation} shows the correlation of the two spectral peak energies for all doubles.
The distribution shows some clustering of the spectral peak values. For the higher of the two peak energies 
$E_2$, the clustering is particularly pronounced close to 7.1~keV and 7.5~keV. The cluster at 7.5~keV comes from
caustic folds amplifying the emission form the receding side of the accretion disk with the positive side pointing 
away from the black hole. In this case, the caustic magnification produces a peak at $E_1$ which changes
continuously when the caustic moves. However, the peak at $E_2$ comes from the unmagnified 
but extremely bright part of the accretion disk approaching the observer at relativistic velocities. 
As this second peak is independent of the caustic magnification, it does not depend on the exact location of the caustic.
The cluster at 7.1~keV comes from  a number of caustics amplifying the bright emission from above the black hole 
(giving a variable $E_1$)  with some magnification of the brightest part of the accretion disk moving $E_2$ 
slightly from 7.5~keV to 7.1~keV.

\begin{figure}
\plotone{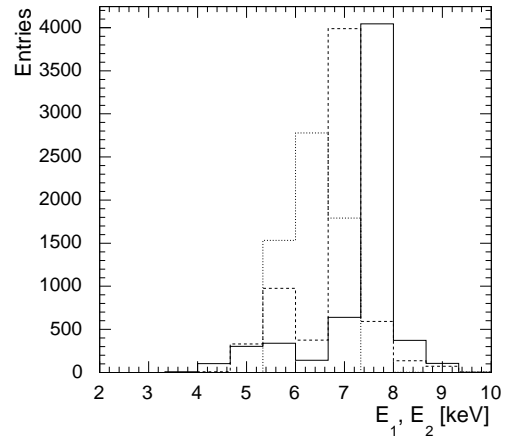}
\caption{\label{fig:h} The distribution of spectral peaks for three lamppost corona heights 
$h\,=\,5\,r_{\rm g}$ (solid line),
$h\,=\,10\,r_{\rm g}$ (dashed line), and
$h\,=\,100\,r_{\rm g}$ (dotted line) 
($a\,=\,0.9$, $i\,=\,82.5^{\circ}$, $\beta\,=\,0.5$).}
\end{figure}

Figure~\ref{fig:h} presents the dependence of the spectral peaks on the lamppost height $h$ for~$a\,=\,0.5$
and $i\,=$~82.5$^{\circ}$. The spectral peak distribution becomes markedly more narrow for 
increasing lamppost height as larger heights reduce the contribution of the inner part of the accretion disk 
with extreme $g$-factors to the observed energy spectra, making a spectral peak close to the emitted 
energy of 6.4 keV more likely.

\begin{figure}
\plotone{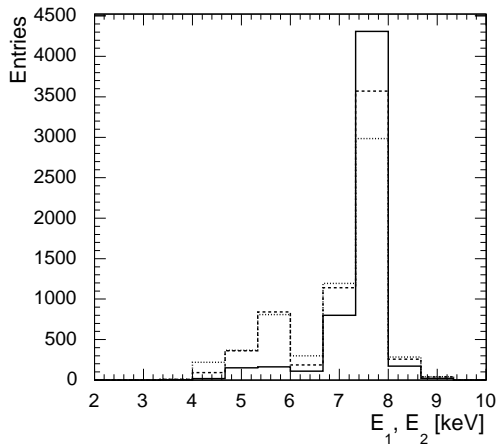}
\plotone{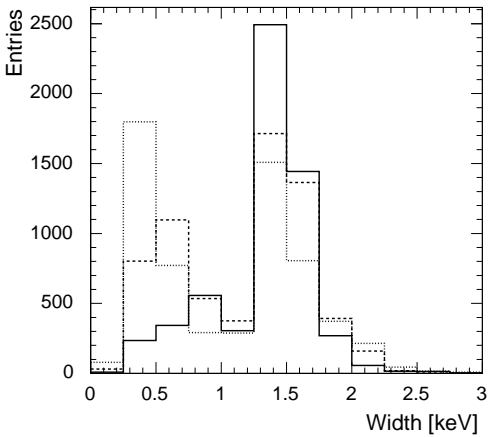}
\caption{\label{fig:beta} The distribution of spectral peaks (upper panel) and spectral peak widths (lower panel)
for $\beta\,=\,0.125$ (solid lines), $\beta\,=\,1$ (dashed lines), and $\beta\,=\,2$ (dotted lines) for $a\,=\,0.5$, $i\,=\,82.5^{\circ}$.
}
\end{figure}

The properties of the spectral peaks depend strongly on the parameter $\beta$. Figure~\ref{fig:beta} 
shows that while the spectral peaks move only to slightly higher energies with increasing $\beta$, the widths of the
spectral peaks increase drastically for larger $\beta$-values. 
The energies increase slightly as higher $\beta$-values make it  more likely that the region of the accretion disk with high magnification 
includes the highly Doppler-boosted emission from the parts of the accretion disk approaching the observer.
The widths increase strongly with increasing $\beta$, as higher $\beta$-values lead to larger regions with substantial 
magnification, resulting in a wider range of $g$-factors contributing to the observed energy spectra.

\begin{figure}
\plotone{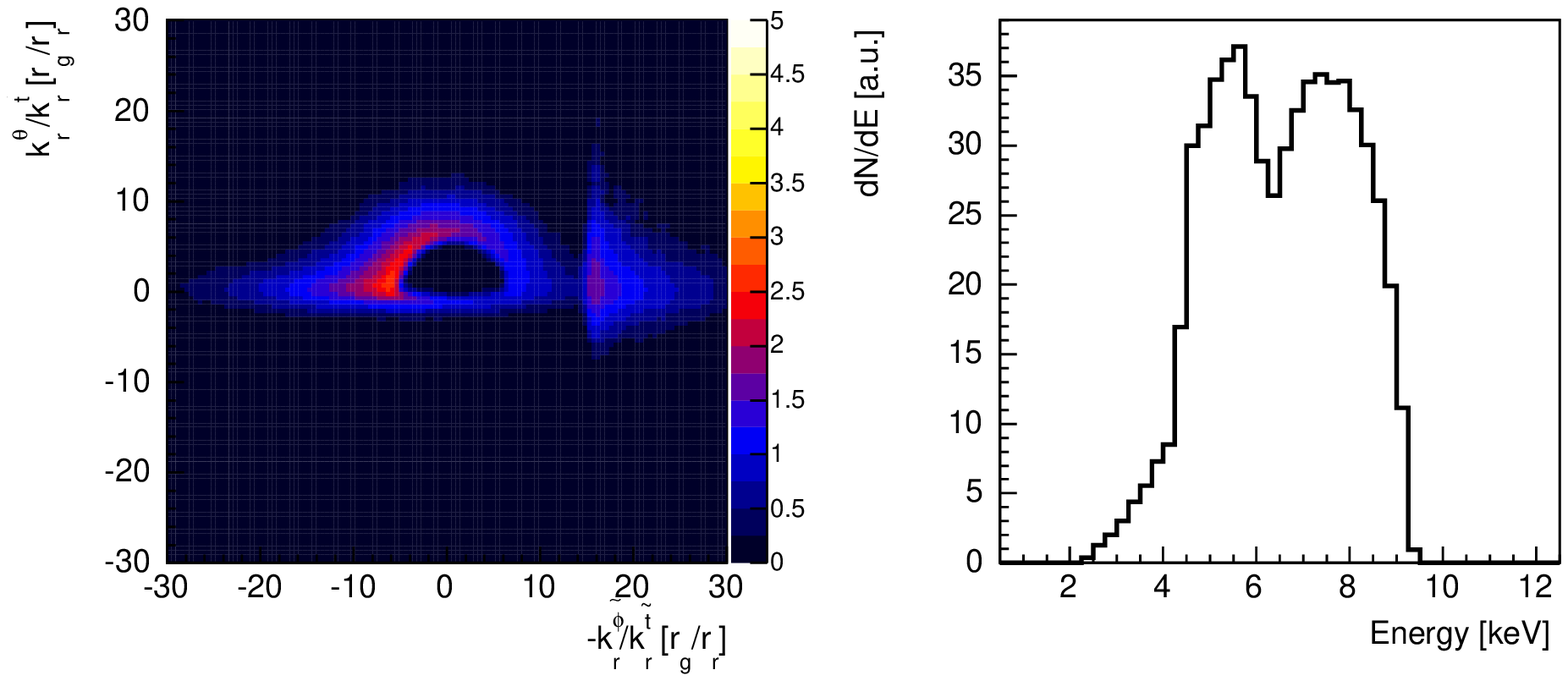}
\plotone{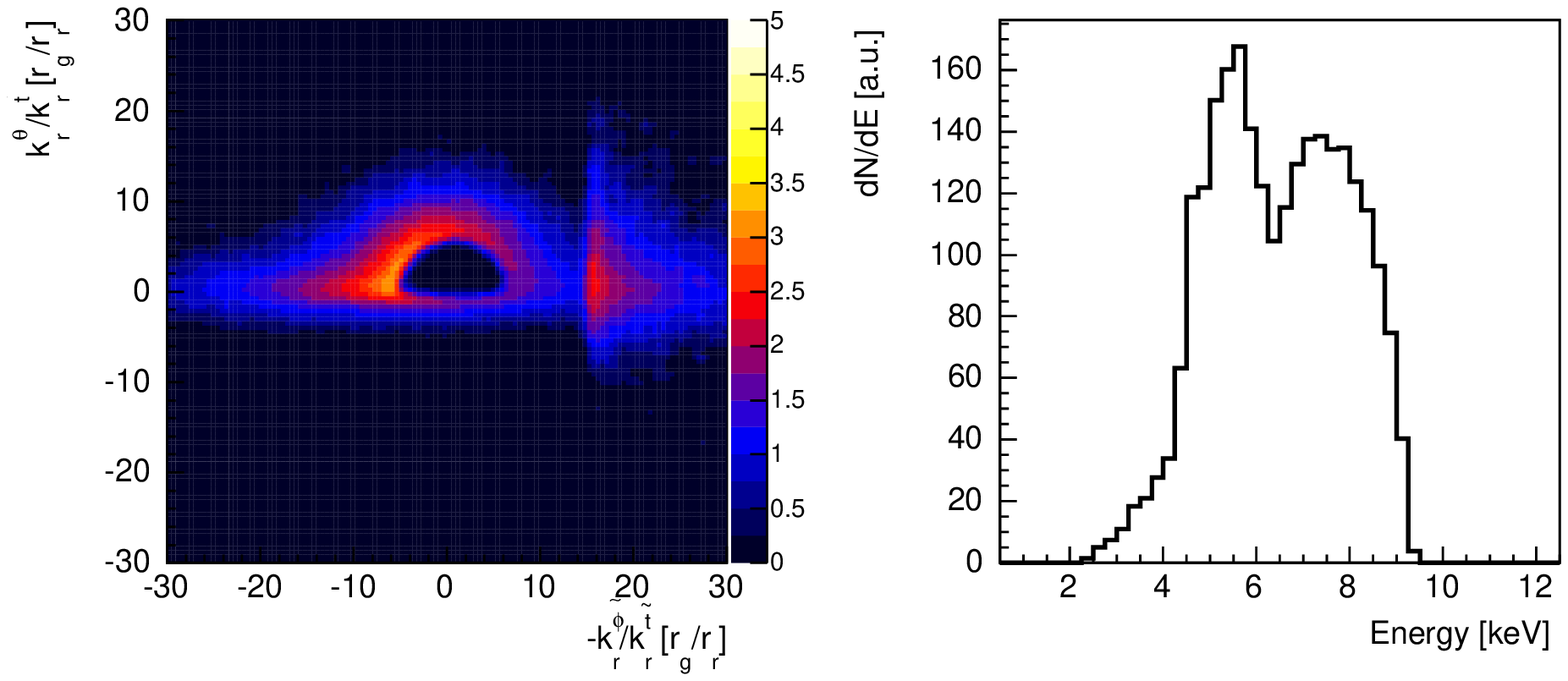}
\caption{\label{fig:pabs} Magnification weighted surface brightness (logarithmic color scale) for the same
configuration as the upper panel of Figure~\ref{fig:m01} but for an increased absorption probability (upper panels, 
$p_{\rm abs}=0.99$ instead of $p_{\rm abs}=0.9$) and for an increased likelihood of scattering rather than
Fe K$\alpha$ emission (lower panels, $R\,=\,5$ rather than $R\,=\,1$).}
\end{figure}

The reflection properties of the accretion disk have a rather small impact on the shape of the energy spectra.
Figure~\ref{fig:pabs} shows the impact of changing the absorption probability from 90\% to 99\% and 
from increasing the  scattering-to-Fe~K$\alpha$ probability ratio from 1:1 to 5:1. 
Comparing the energy spectrum of the upper panel of Figure~\ref{fig:m01} with those of Figure~\ref{fig:pabs}
shows that the energies and widths of the spectral peaks do not change substantially.
\section{Determination of the Black Hole, Accretion Disk, Corona, and Lens Parameters}
\label{ML}
In this section we investigate if the parameters describing the spectral peaks contain enough information
to allow us to to constrain several black hole, accretion disk, 
corona and lens parameters $P\,=\left\{a,i,h,\beta\right\}$.
We do so in two ways: first, we check if the detection of a few extreme line properties can 
be used to constrain the allowed range of certain parameters.  
Second, we assume a certain fiducial parameter combination, generate simulated data sets, 
fit these simulated data sets with a maximum likelihood fit, and assess how well the fit
recovers the input parameters.   

The first type of analysis gives useful constraints on the inclination and corona height. 
\begin{figure}
\plotone{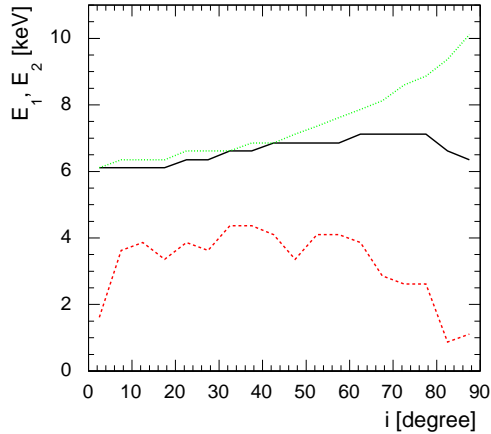}
\plotone{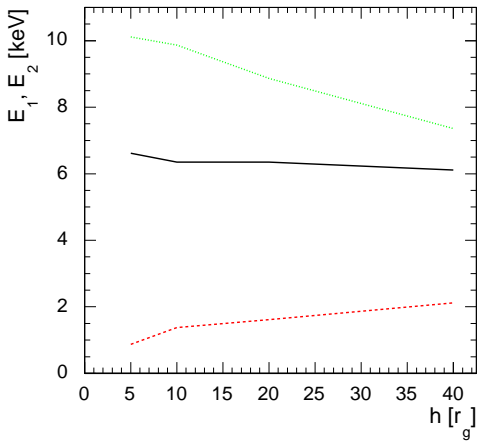}
\caption{\label{fig:eRange} Range of the simulated spectral peak energies as a function of the inclination $i$ 
(upper panel) and corona height $h$ (lower panel).The dashed, solid, and dotted lines show the minimum, 
median, and maximum spectral peak values. The minimum, median, and maximum values were determined by
analyzing the distribution of the spectral peaks found in the energy spectra for all simulated black hole spins, 
all simulated caustic positions and orientations, all simulated corona heights (upper panel), 
and all possible inclinations (lower panel). Energy spectra with a single peak at energy $E_1$ 
contribute a single entry to the peak energy distribution.  Energy spectra with two peaks 
at energies $E_1$ and $E_2$ contribute two entries to the peak energy distribution. 
}\vspace*{2ex}
\end{figure}
Figure \ref{fig:eRange} shows the range of the simulated spectral peaks as a function of inclination (upper panel)
and corona height (lower panel). The minimum and maximum spectral peaks of 
3.9~keV and $\sim$7.4~keV detected for RX~J1131$-$1231 (C17) require an inclination $i\,\ge\,60^{\circ}$
and a corona height of $h\,\le$~40~$r_{\rm g}$. 
\begin{figure}
\plotone{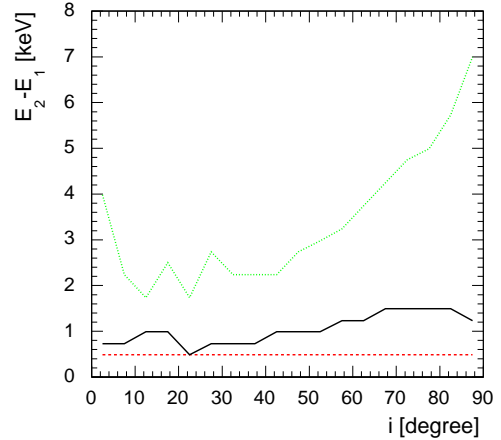}
\plotone{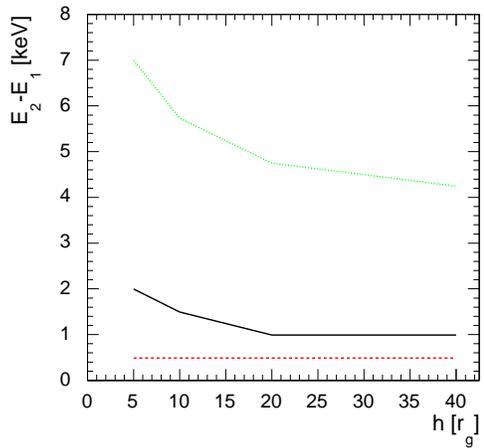}
\caption{\label{fig:deRange} Range of the differences of the spectral peak energies of all doubles as function 
of the inclination $i$ (upper panel) and corona height $h$ (lower panel). 
The dashed, solid, and dotted lines show the minimum, median, and maximum values.}
\end{figure}
Figure \ref{fig:deRange} shows in a similar way the ranges of the differences of the spectral peak energies of
all doubles as a function of inclination (upper panel) and corona height (lower panel). 
Explaining differences of 4.5~keV such as those detected for RX~J1131$-$1231 
with this paradigm would imply inclinations of $i\,\ge\,70^{\circ}$ and corona heights 
of $h\,\le\,30$~$r_{\rm g}$.

The comparison of the simulated energy separations of all doubles (Figure 7) 
with the ones observed for RX J1131$-$1231 (Figure 20, C17) indicates 
a high black hole spin. However, we postpone deriving firm constraints until
we accurately modeled the {\it Chandra} detection biases.

\begin{figure}
\begin{center}
\vspace*{-2ex}
\includegraphics[width=0.32\textwidth]{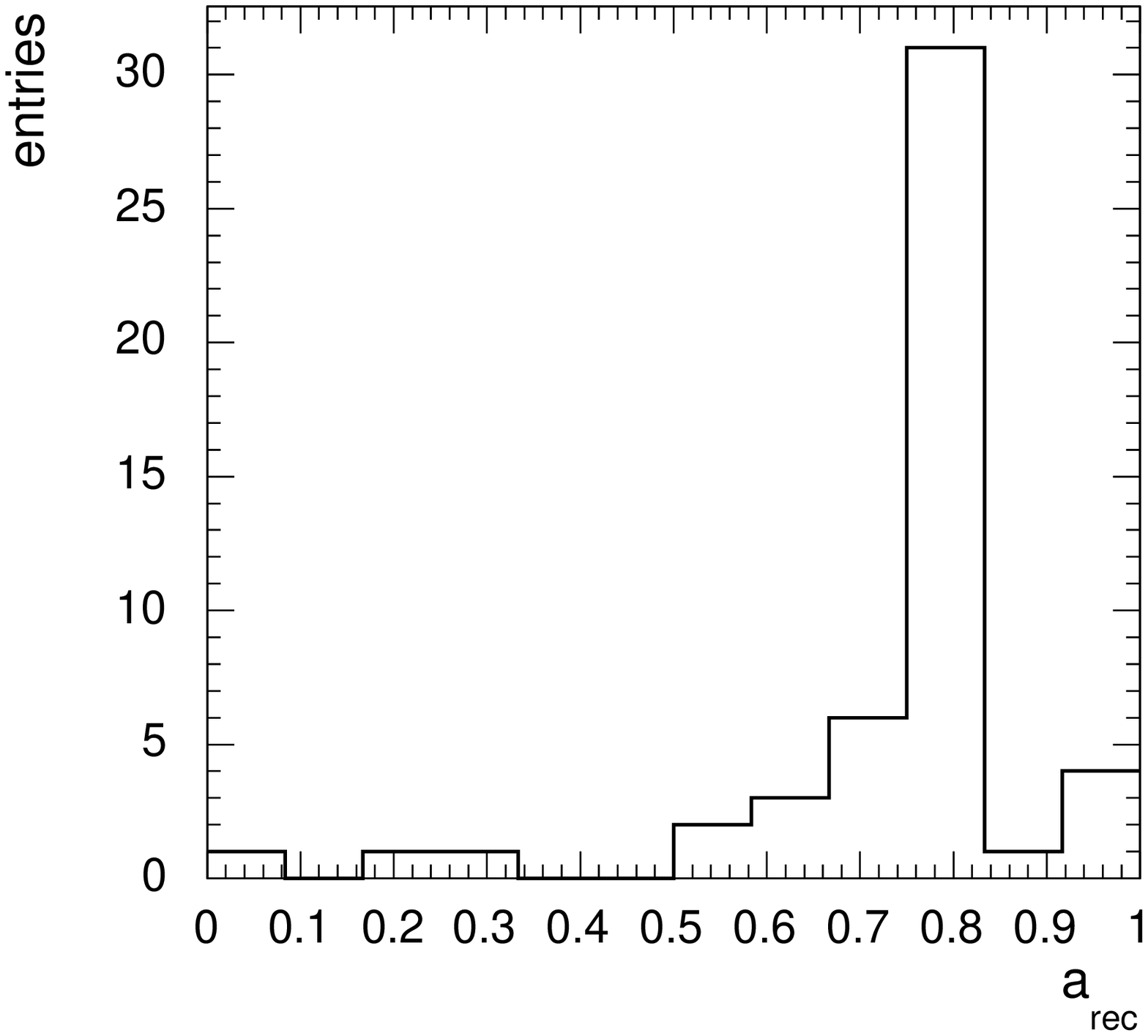}
\vspace*{-4.5ex}
\includegraphics[width=0.32\textwidth]{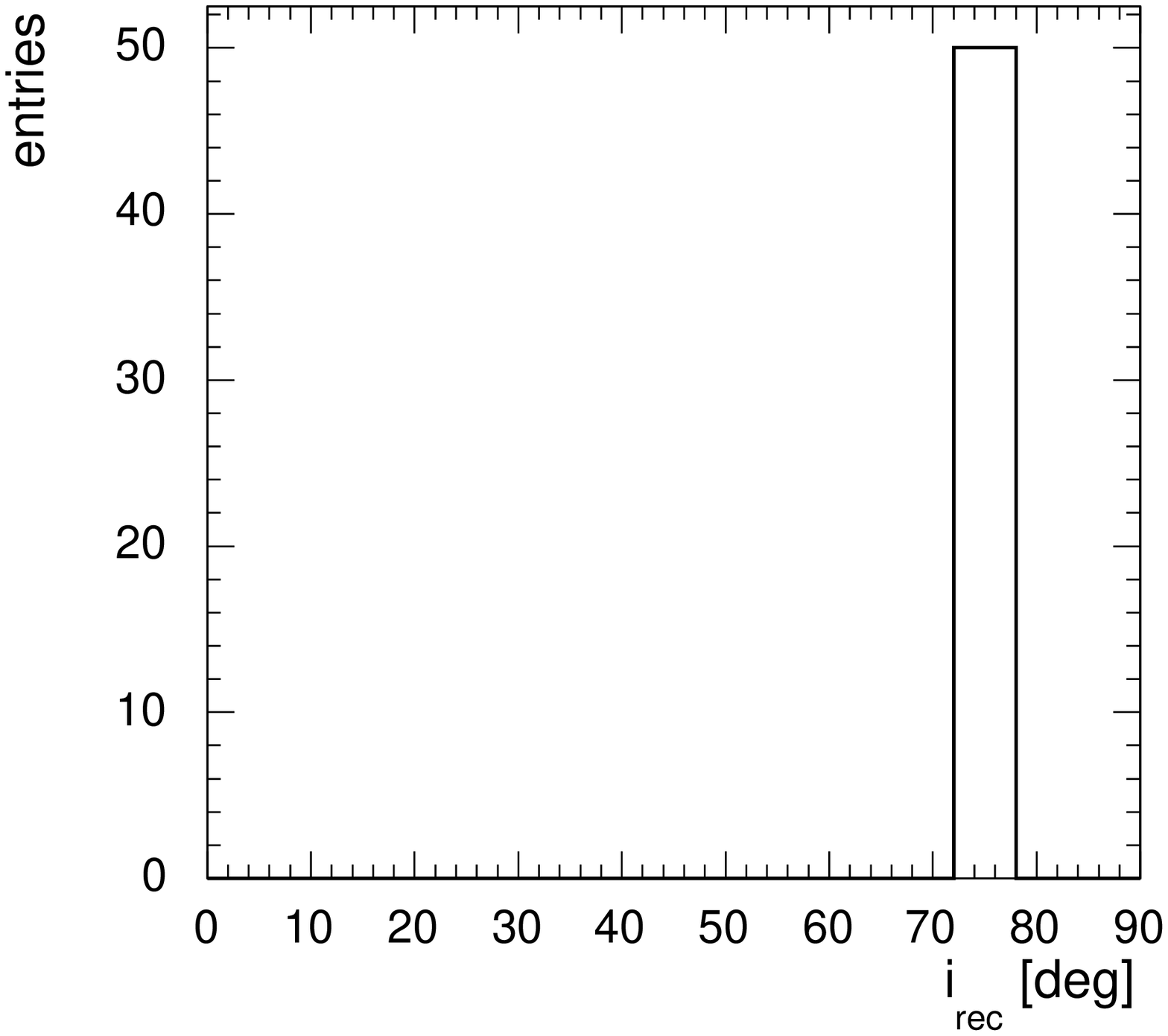}
\vspace*{-4.5ex}
\includegraphics[width=0.32\textwidth]{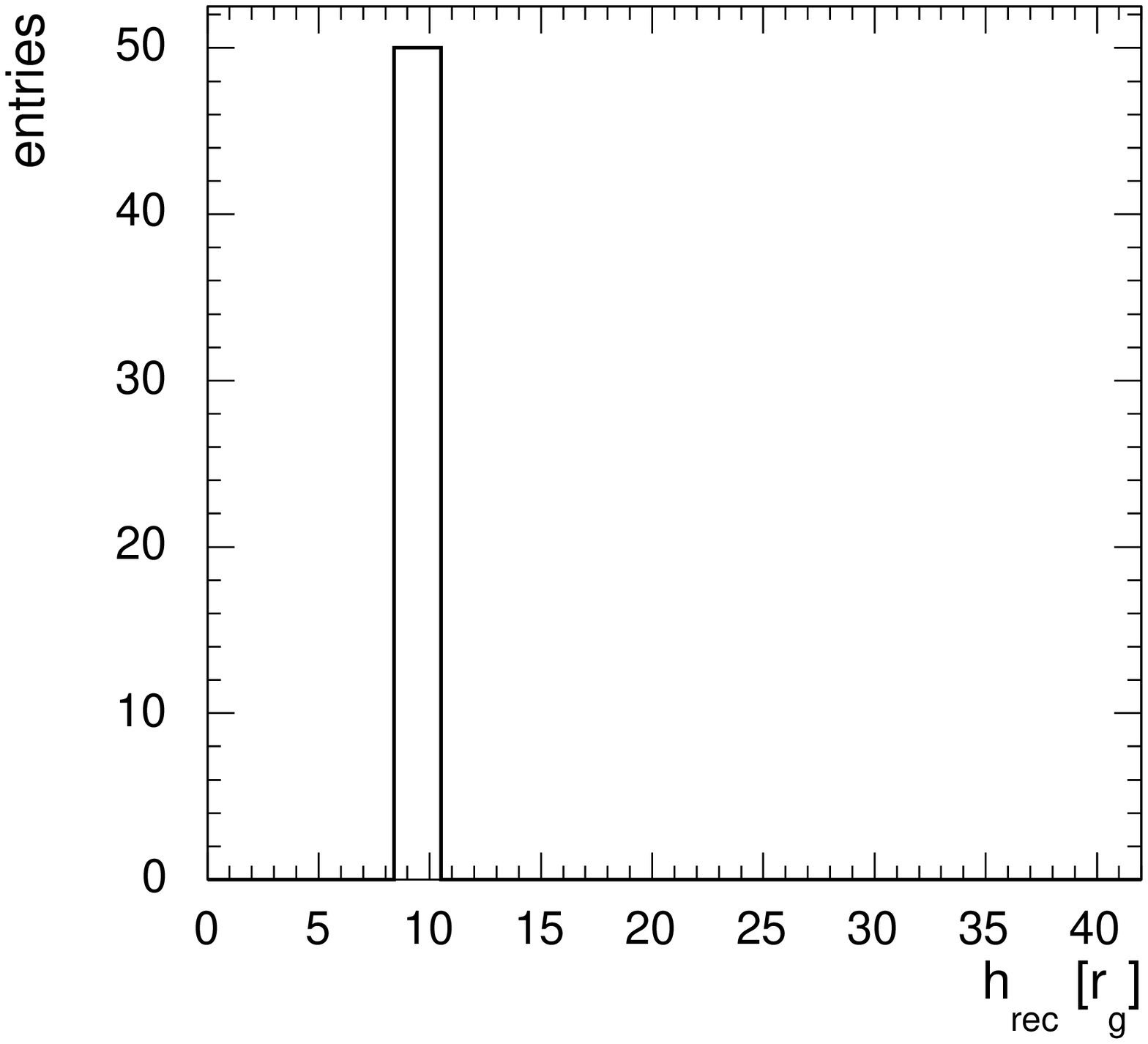}
\vspace*{-4.5ex}
\includegraphics[width=0.32\textwidth]{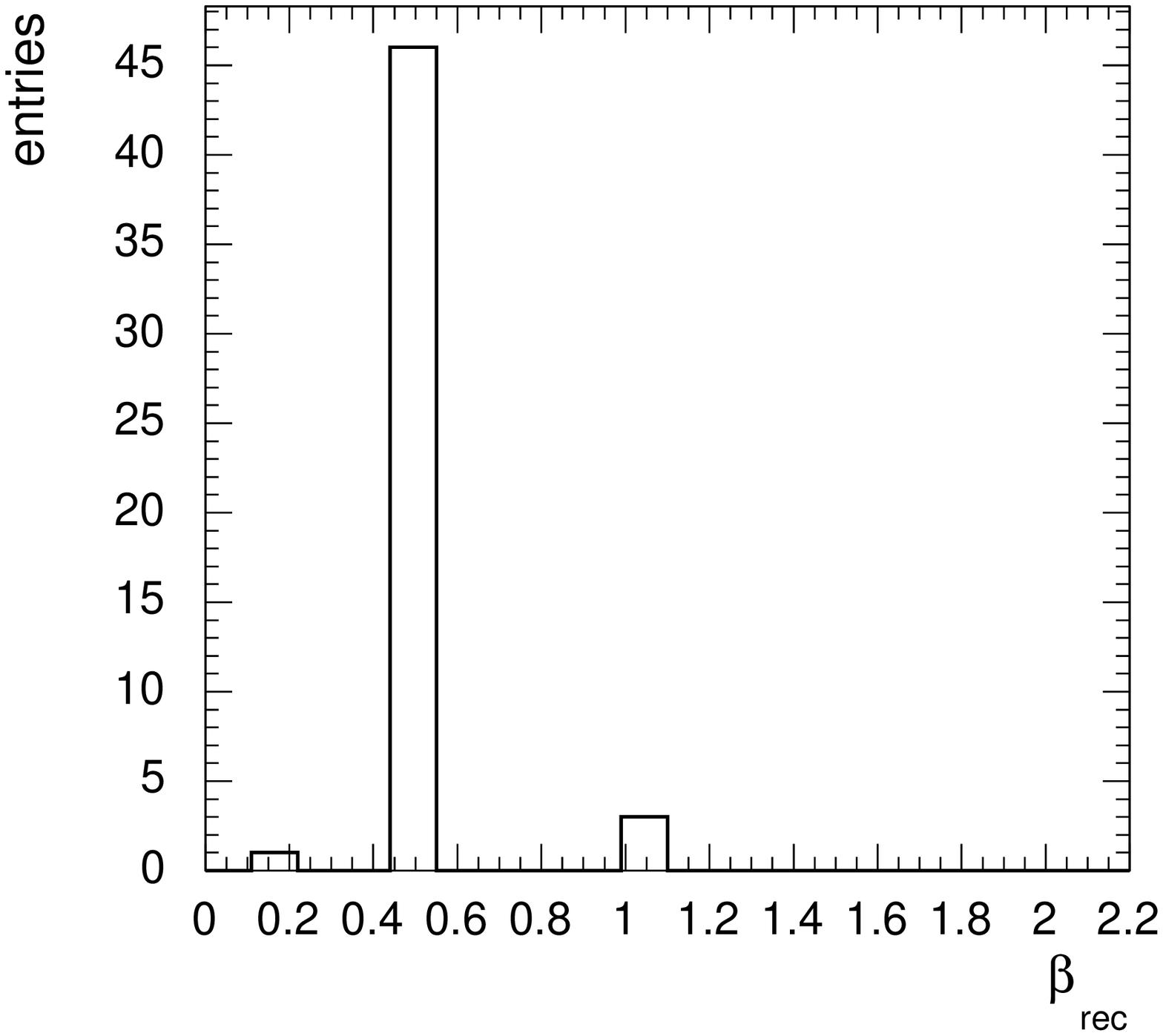}
\vspace*{-2.5ex}
\end{center}
\caption{\label{fig:fit} Distribution of (from top to bottom) reconstructed black hole spin, 
accretion disk inclination, corona height, and $\beta$-parameter
for simulated data sets with 40 measured microlensed energy spectra
for the input values  $P_0\,=$ $\left\{a_0=0.8,i=77.5^{\circ},h=10\,r_{\rm g},\beta=0.5\right\}$.
}
\end{figure}

For the second approach we choose a fiducial parameter configuration $P_0\,=\,\left\{a_0,i_0,h_0,\beta_0\right\}$
and generate $N$ data sets by randomly drawing $L$ energy spectra simulated for the configuration 
$P_0$ and adding normally distributed errors to the parameters.
We base the maximum likelihood fit on the spectral peak energy $E_1$ and width $w_1$
in the case of single-peaked energy spectra, and the energies $E_1$, $E_2$, and widths $w_1$, $w_2$, 
and the ratio of the normalization constants $\rho_n\,=\,n_2/n_1$ in the case of double-peaked energy spectra. 
The rationale of using $\rho_n$ rather than $n_1$ and $n_2$ is that the former is independent of the 
time-variable inherent source variability, while the latter are not. We assume Gaussian widths  
$\sigma_E / E\,=$~ 0.02,  $\sigma_w / w\,=$~0.3, $\sigma_{\rho_n}/\rho_n \,=$~0.03 
similar to the errors of the {\it Chandra} data set.

Each of the $N$ data set is fitted by finding the parameter combination $P\,=\,P^*$ maximizing the likelihood function:
\begin{equation}
\Lambda(P)\,=\,
\prod\limits_{\rm l=1}^{L} p_l(P)
\end{equation}
with $p_l$ being the probability for detecting the parameters of the $l^{\rm th}$ energy spectrum:
\begin{equation}
 p_l(P)\,=\,\frac{1}{M} \sum_{m=1}^{M} q_{l,m}(P).
 \end{equation}
The sum over $m$ runs over all $M$ simulated template energy spectra generated for configuration $P$
for different caustic crossing angles and offsets, and $q_{l,m}$ is the probability that template $m$ produces the result $l$. 
If the numbers of spectral peaks of energy spectra $l$ and $m$ do not agree, $q_{l,m}\,=\,0$. 
If they do agree, $q_{l,m}$ equals the product of the Gaussian probability density functions 
of all the relevant parameters. 
We find that the method recovers the input parameters with good accuracy. 
Figure~\ref{fig:fit} shows the distribution of the best-fit parameters $P^*$ for 
$P_0\,=$ $\left\{a_0=0.8,i=77.5^{\circ},h=10\,r_{\rm g},\beta=0.5\right\}$ for 
50 simulated data sets with 40 line detections each. 
The accuracy of the reconstruction improves with the number of lines used for the analysis.
Simulating 50 data sets with 200 line detections each, we find that the fit recovers 
the fiducial parameter values for all 50 data sets.
Fitting the actual data will require more detailed simulations of the microlensing 
(including an exploration of the microlensing parameter space), and detailed modeling
of the energy dependent detection biases resulting from the continuum emission and 
the associated statistical noise, and the  {\it Chandra} effective area, energy resolution, 
image cross talk, and background.
\vspace*{2ex}
\section{Summary and Discussion}
\label{summary}
In this paper, we study the observational signatures from the microlensing of the Fe~K$\alpha$ 
emission from QSOs. Introducing the concept of a virtual source embedded 
in flat spacetime mimicking the effects of the source embedded in the Kerr spacetime 
allows us to divide the problem into (i) raytracing simulations in the Kerr spacetime and 
(ii) the simulations of the effect of microlensing. 
The amplification close to caustic folds selectively amplifies the emission from a slice of the 
accretion disk.

The microlensing of the Fe~K$\alpha$  emission can produce single peaked and double-peaked Fe~K$\alpha$ 
energy spectra with variable peak energies. 
We generate simulated Fe~K$\alpha$ energy spectra which can be compared to observed ones. 
We find that the range of spectral peak energies depends strongly on the inclination of the accretion disk
and the lamppost height. The energy difference between two spectral peaks depends strongly
on the black hole spin. Using simple $r$-independent parameterizations of the photon absorption, 
scattering and Fe~K$\alpha$ emission probabilities, we find that the shape of the 
Fe~K$\alpha$ energy spectra do not depend strongly on the particular parameter choices. 
The detection of extreme spectral peak centroids and spectral peak energy separations 
can constrain the accretion disk inclination and the height of the lamppost corona.
In the case of RX~J1131$-$1231, the detected difference of double line detections
can be explained for inclinations $i \ge 70^{\circ}$ and corona heights $h\le 30\,r_{\rm g}$.
We present a maximum likelihood fit that uses the energy centroids and widths of all  
spectral peaks, and, in the case of energy spectra with two or more spectral peaks, the ratio of the 
spectral peak normalizations as input. Under the simplifying assumptions made in this paper, 
the fit can recover the black hole spin, 
accretion disk inclination, the corona height, and the microlensing parameter $\beta$.

The simulations of this paper assume a simple lamppost corona geometry.  We plan to study the impact 
of the observable signatures on the corona properties in an upcoming paper. 
The most severe limitation of our study is the use of a simple parameterization of the magnification
close to caustics with a constant magnification scale factor. The treatment does not capture 
the source plane density of caustics, the statistical distribution of the magnification scale factors, and
correlations between the results of different observations.  
A more detailed analysis will need to account for the convergence and shear from the lensing galaxy and for the surface density and mass function of the microlensing stars. 
Describing the observed data may require to describe the 
correlations between different observations by considering trajectories 
of the source across the magnification maps.  
Fitting actual data will require to model the probability that {\it Chandra} detects the simulated spectral peaks.
The latter probability depends on the line and continuum fluxes, the exposure time of the observations, 
the energy and width of the shifted lines and {\it Chandra's} energy dependent effective area, energy resolution,
and image cross talk. A full analysis would add the simulated Fe~K$\alpha$ emission to a 
model of the continuum emission. After convolving the resulting energy spectra with the {\it Chandra} 
instrument response, the simulated data sets can be analyzed in exactly the same way as the observed data sets.
One of the payoffs of such a detailed analysis would be to tighten the constraint on the spin of the black hole of 
the quasar  RX~J1131$-$1231 of $a \simgt 0.8$ based on the analysis of the observed distribution of the 
energy difference of energy spectra with two spectral peaks (C17).

Microlensing of the Fe~K$\alpha$ emission from quasars should occur at some level.
The simulations presented in this paper show that precision measurements of the line shapes 
enable tomographic studies of the accretion disk and the black hole spacetime. 
Compared to the analysis of the shape of the Fe~K$\alpha$ emission from unlensed 
nearby narrow line type I Seyferts, studies of microlensed sources have the advantage that the 
detection of spectral peaks is less susceptible to details of the continuum subtraction than the 
analysis of the extreme wings of the Fe~K$\alpha$ line. 
\begin{acknowledgments}
HK would like to thank NASA (grant \#NNX14AD19G) and the 
Washington University McDonnell Center for the Space Sciences for financial support.
GC would like to acknowledge financial support from NASA via the Smithsonian Institution 
grants SAO GO4-15112X, GO3-14110A/B/C, GO2-13132C, GO1-12139C, and GO0-11121C.
HK thanks Fabian Kislat for fruitful discussions.
\end{acknowledgments}

\end{document}